\documentclass[12pt]{article}

\usepackage{amsmath,amsfonts,amssymb,bbm,latexsym}
\usepackage[mathscr]{eucal}

\def\PCPqed{\hspace*{\fill}\ensuremath{\Box}}
\newcommand{\smin}{\,\raisebox{0.06em}{${\scriptstyle \in}$}\,}

\newcommand{\ssmin}{\,\raisebox{0.06em}{${\scriptscriptstyle \in}$}\,}

\newcommand{\smwedge}{{\scriptstyle \wedge}}

\newcommand{\smcirc}{\,\raisebox{0.1ex}{${\scriptstyle \circ}$}\,}
\newcommand{\bwedge}{\raisebox{0.2ex}{${\textstyle \bigwedge}$}}
\newcommand{\upoast}{\ensuremath{%
 ^{\raisebox{0.1ex}{$\scriptscriptstyle \bigcirc$} \hspace{-0.5em} \ast}}}
\newcommand{\upostar}{\ensuremath{%
 ^{\raisebox{0.05ex}{$\scriptscriptstyle \bigcirc$} \hspace{-0.5em} \star}}}
\newcommand{\smupoast}{\ensuremath{%
 ^{\raisebox{-0.25ex}{$\textstyle \circ$} \hspace{-0.42em}
  {\scriptscriptstyle \ast}}}}
\newcommand{\smupostar}{\ensuremath{%
 ^{\raisebox{-0.25ex}{$\textstyle \circ$} \hspace{-0.42em}
   \raisebox{0.05ex}{$\scriptscriptstyle \star$}}}}

\newtheorem{prp}{Proposition}[section]
\newtheorem{lem}[prp]{Lemma}
\newtheorem{thm}[prp]{Theorem}
\newtheorem{dfn}[prp]{Definition}

\begin{document}

\title{The Poisson Bracket for Poisson Forms \\
       in Multisymplectic Field Theory}
\author{
  Michael Forger$\,^1\,$\thanks{Partially supported by CNPq, Brazil}~,
  Cornelius Pauf\/ler$\,^2$~
  and
  Hartmann R\"omer$\,^3\,$\thanks{Partially supported by FAPESP, Brazil}}
\date{\normalsize
      $^1\,$ Departamento de Matem\'atica Aplicada,~~\mbox{} \\
      Instituto de Matem\'atica e Estat\'{\i}stica, \\
      Universidade de S\~ao Paulo, \\
      Caixa Postal 66281, \\
      BR--05311-970~ S\~ao Paulo, S.P., Brazil \\[4mm]
      $^{2,3}\,$ Fakult\"at f\"ur Physik \quad \mbox{} \\
      Albert-Ludwigs-Universit\"at Freiburg im Breisgau \\
      Hermann-Herder-Stra\ss e 3 \\
      D--79104~ Freiburg i.Br., Germany 
      }

\footnotetext[1]{\emph{E-mail address:} \textsf{forger@ime.usp.br}}
\footnotetext[2]{\emph{E-mail address:}
 \textsf{cornelius.paufler@physik.uni-freiburg.de}}
\footnotetext[3]{\emph{E-mail address:}
 \textsf{hartmann.roemer@physik.uni-freiburg.de}}
\maketitle

\thispagestyle{empty}

\begin{abstract}
\noindent
We present a general definition of the Poisson bracket between differential
forms on the extended multiphase space appearing in the geometric formulation
of first order classical field theories and, more generally, on exact multi%
symplectic manifolds. It is well defined for a certain class of differential
forms that we propose to call Poisson forms and turns the space of Poisson
forms into a Lie superalgebra.
\end{abstract}
\vspace{-5mm}
\begin{flushright}
 \parbox{12em}
 {\begin{center}
   Universit\"at Freiburg \\
   THEP 02/03 \\
   Universidade de S\~ao Paulo \\
   RT-MAP\,-\,0201 \\
   February 2002
  \end{center}}
\end{flushright}

\newpage

\section{Introduction}

The multiphase space approach to classical field theory, whose origins
can be traced back to the early work of Hermann Weyl on the calculus of
variations, has recently under\-gone a rapid development, but a number
of conceptual questions is still open.

The basic idea behind all attempts to extend the covariant formulation
of classical field theory from the Lagrangian to the Hamiltonian domain
is to treat spatial derivatives on the same footing as time derivatives.
This requires associating to each field component $\varphi^{\,i}$ not
just its standard canonically conjugate momentum $\pi_i^{}$ but rather
$n$ conjugate momenta $\pi\:\!_i^\mu$, where $n$ is the dimension of
space-time. If one starts out from a~Lagrangian $\mathcal{L}$ depending
on the field and its first partial derivatives, these are obtained by
the covariant Legendre transformation
\[
 \pi_i^\mu~
 =~\frac{\partial \mathcal{L}}{\partial \, \partial_\mu \varphi^{\,i}}~.
\]
This allows to rewrite the standard Euler-Lagrange equations of field theory,
\[
 \partial_\mu \,
 \frac{\partial \mathcal{L}}{\partial \, \partial_\mu \varphi^i} \, - \,
 \frac{\partial \mathcal{L}}{\partial \varphi^i}~=~0
\]
as a covariant first order system, the covariant Hamiltonian equations or
De Donder - Weyl equations
\[
 \frac{\partial \mathcal{H}}{\partial \pi_i^\mu}~
 =~\partial_\mu \varphi^i \quad , \quad
 \frac{\partial \mathcal{H}}{\partial \varphi^i}~
 = \;- \, \partial_\mu \pi_i^\mu
\]
where
\[
 \mathcal{H}~=~\pi_i^\mu \, \partial_\mu \varphi^i \, - \, \mathcal{L}
\]
is the covariant Hamiltonian density or De Donder - Weyl Hamiltonian.

Multiphase space (ordinary as well as extended) is the geometric environment
built by appropriately patching together local coordinate systems of the
form $(q^i,p\>\!_i^\mu)$~-- instead of the canonically conjugate variables
$(q^i,p_i)$ of mechanics~-- together with space-time coordinates $x^\mu$
and, in the extended version, a further energy type variable that we shall
denote by $p$ (without any index). In the recent literature on the subject,
particular attention has been devoted to the so-called multisymplectic
form, which is naturally defined on extended multiphase space and is the
geometric object replacing the symplectic form $\, \omega = dq^i \,\smwedge\;
dp_i^{} \,$ of mechanics.

The advantage of this approach as compared to the orthodox strategy of
treating field theoretical models as infinite-dimensional dynamical
systems is threefold. First, general covariance (and in particular,
Lorentz covariance) is trivially achieved. Second, by working on
multiphase space which is a finite-dimensional manifold, one
automatically avoids all the functional analytic complications
that plague the orthodox method. Third, space-time locality is
also automatically guaranteed, since one works with the field
variables and their first derivatives or conjugates of these at
single points of space-time, rather than with fields defined over
entire hypersurfaces: integration is deferred to the very last step
of every procedure. Of course, there is also a price to be paid for
all these benefits, namely that the obvious duality of classical
mechanics between coordinates and momenta is lost. As a result,
there is no evident multiphase space quantization procedure.
What seems to be needed is a new and more sophisticated concept
of ``multi-duality'' to replace the standard duality underlying
the canonical commutation relations.

Certainly, an important step towards a better understanding of what might
be the nature of this ``multi-duality'' and that of a multiphase space
quantization procedure is the construction of Poisson brackets within
this formalism. After all, the Poisson bracket should be the classical
limit of the commutator of quantum theory. Surprisingly, this is to a
large extent still an open problem. Our approach to the question has
been motivated by the work of Kanatchikov \cite{Ka1,Ka2}, who seems
to have been the first to propose a Poisson bracket between differential
forms of arbitrary degree in multimomentum variables and to analyze the
restrictions that must be imposed on these forms in order to make this
bracket well-defined: he uses the term ``Hamiltonian form'' in this
context, although the concept as such is of course much older. It must
be pointed out, however, that Kanatchikov's approach is essentially local
and makes extensive use of features that have no invariant geometric meaning,
such as a systematic splitting into horizontal and vertical parts; moreover,
his definition of Hamiltonian forms is too restrictive. We avoid all these
problems by working exclusively within the multisymplectic framework and
on the extended multiphase space, instead of the ordinary one: this leads
naturally to a definition of the concept of a Poisson form which is more
general than Kanatchikov's notion of a Hamiltonian form, as well as to a
coordinate-independent definition of the Poisson bracket between any two
such forms. In fact, most of the concepts involved do not even depend on
the explicit construction of extended multiphase space but only on its
structure as an exact multisymplectic manifold, and we shall make use
of this fact in order to simplify the treatment whenever possible.

The paper is organized as follows. In Sect.\ 2, we give a brief review of
some salient features of the multiphase space approach to the geometric
formulation of first \mbox{order} classical field theories, following Ref.\
\cite{CCI} and, in particular, Ref.\ \cite{GIM}, to which the reader
is referred for more details and for the discussion of many relevant
examples; this material is included here mainly in order to fix notation
and make our presentation reasonably self-contained. The main point is
to show that the extended multiphase space of field theory does carry
the structure of an exact multisymplectic manifold (in fact it seems
to be the only known example of a multisymplectic manifold). \linebreak
In Sect.~3, we introduce the concept of a Poisson form on a general
multisymplectic manifold, specify the notion of an exact multisymplectic
manifold, define a Poisson bracket between Poisson forms on exact multi%
symplectic manifolds and prove our main theorem, which states that this
bracket satisfies the usual axioms of a Lie superalgebra. The construction
generalizes the corresponding one for Hamiltonian $(n-1)$-forms on the
extended multiphase space of field theory given by two of the present
authors in a previous paper \cite{FR1}: the idea is to modify the
standard formula that had been adopted for decades [6-11], even though
it fails to satisfy the Jacobi identity, by adding a judiciously chosen
exact form that turns out to cure the defect. Here, we show that the
same trick works for forms of arbitrary degree, provided one introduces
appropriate sign factors. In both cases, it is the structure of the
correction term that requires the underlying manifold to be exact
multisymplectic and not just multisymplectic. \linebreak In Sect.~4,
we define the notion of an exact Hamiltonian multivector field on an
exact multisymplectic manifold and show that by contraction with the
multicanonical form $\theta$, any such multivector field gives rise
to a Poisson form; moreover, this simple prescription yields an anti%
homomorphism of Lie algebras (with respect to the standard Schouten bracket
of multivector fields and the Poisson bracket of Poisson forms introduced
here). \linebreak It can be viewed as an extension, from vector fields
to multivector fields, of the universal part of the covariant momentum
map~\cite{GIM}, which is the geometric version of the construction of
Noether currents and the energy-momentum tensor in field theory, and
we shall therefore refer to it as the universal multimomentum map.
In Sect.~5, we return to the case of extended multiphase space and
discuss other examples for the construction of Poisson forms. More
specifically, we show that arbitrary functions are Poisson forms
(of degree $0$) and find that Kanatchikov's Hamiltonian forms, when
pulled back from ordinary to extended multiphase space by means of the
appropriate projection, constitute a special class of Poisson forms.
The complete determination of the space of Poisson forms of arbitrary
degree $>0$ on extended multiphase space, together with that of exact
Hamiltonian and locally Hamiltonian multivector fields of arbitrary
degree $<n$, is a technically demanding problem whose solution will
be presented separately~\cite{FPR2}. The paper concludes with two
appendices: the first presents a number of important formulas from
the multivector calculus on manifolds, related to the definition and
main properties of the Schouten bracket and the Lie derivative of
differential forms along multivector fields, while the second shows
how, given a connection in a fiber bundle, one can construct induced
connections in various other fiber bundles derived from it, including
the multiphase spaces of geometric field theory; this possibility is
important for the comparison of the multisymplectic formalism with
other approaches that have been proposed in the literature and to a
certain extent depend on the \emph{a priori\/} choice of a connection.

Finally, we would like to point out that there exists another construction
of a covariant Poisson bracket in classical field theory, based on the same
functional approach that underlies the construction of ``covariant phase
space'' of Crnkovic-Witten~\cite{CW,Cr} and Zuckerman~\cite{Zu}. This
bracket, originally due to Peierls \cite{Pe} and further ela\-borated
by de Witt~\cite{DW1,DW2}, has recently been adapted to the multiphase
space approach by Romero~\cite{Ro} and shown to be precisely the Poisson
bracket associated with the symplectic form on covariant phase space
introduced in Refs \cite{CW,Cr} and \cite{Zu}; these results will be
presented elsewhere \cite{FR2}. It would be interesting to identify
the relation between that bracket and the one introduced here;
this question is presently under investigation.


\section{Multiphase spaces in geometric field theory}

The starting point for the geometric formulation of classical field theory
is the choice of a \emph{configuration bundle}, which in general will be a
fiber bundle over space-time whose sections are the fields of the theory
under consideration. In what follows, we shall denote its total space by
$E$, its base space by $M$, its typical fiber by $Q$ and the projection
from $E$ to $M$ by $\pi$; the dimensions are
\begin{equation}
 \dim\, M~=~n~~,~~\dim\, Q~=~N~~,~~\dim\, E~=~n+N~.
\end{equation}
In field theoretical models, $M$ is interpreted as {\em space-time} whereas
$Q$ is the \emph{configuration space} of the theory~-- a manifold whose (local)
coordinates describe internal degrees of freedom.\footnote{This interpretation
changes in the theory of strings and membranes.} The total space $E$ is
locally but not necessarily globally isomorphic to the Cartesian product
$\, M \times Q$, but it must be stressed that even when the configuration
bundle is globally trivial, there will in general not exist any preferred
trivialization, and it is precisely the freedom to change trivialization
that allows to incorporate gauge theories into the picture. Another point
that deserves to be emphasized is that the configuration bundle does in
general not carry any additional structures: these only appear when one
focusses on special classes of field theories.
\begin{itemize}
 \item \emph{Vector bundles} arise naturally in theories with \emph{linear
       matter fields} and also in general relativity: the metric tensor is an
       example.
 \item \emph{Affine bundles} can be employed to incorporate \emph{gauge
       fields}, since connections in a principal $G$-bundle $P$ over
       space-time $M$ can be viewed as sections of the \emph{connection
       bundle} of $P$~-- an affine bundle $CP$ over $M$ constructed from
       $P$.
 \item \emph{General fiber bundles} are used to handle \emph{nonlinear
       matter fields}, in particular those corresponding to maps from
       space-time $M$ to some target manifold $Q$: a standard example
       are the nonlinear sigma models.
\end{itemize}
In order to cover this variety of situations, the general constructions
on which the \mbox{geometric} formulation of classical field theory is
based must not depend on the choice of any additional structure on the
configuration bundle. This requirement is naturally satisfied in the
multiphase space formalism~-- in contrast to the majority of similar
\mbox{approaches} that have over the last few decades found their way
into the literature: most of these depend on the \emph{a priori\/} choice
of a connection in the configuration bundle, thus excluding gauge theories
in which connections must be treated as dynamical variables and not as
fixed background fields.

The multiphase space approach to first order classical field theory follows
the same general pattern as the standard formalism of classical mechanics on
the tangent and cotangent bundle of a configuration space $Q$ \cite{AM,Ar}.%
\footnote{The term ``first order'' refers to the fact that the Lagrangian
is supposed to be a pointwise defined function of the coordinates or fields
and of their derivatives or partial derivatives of no more than first order;
higher order derivatives should be eliminated, e.g., by introducing
appropriate auxiliary variables.} However, the correspondence between
the objects and concepts underlying the geometric formulation of mechanics
and that of field theory becomes fully apparent only when one reformulates
mechanics so as to incorporate the time dimension. (This is standard practice,
e.g., in the study of non-autonomous systems, that is, mechanical systems
whose Lagrangian\,/\,Hamiltonian depends explicitly on time, such as systems
of particles in time-dependent external fields. Additional motivation is
provided by relativistic mechanics where Newton's concept of absolute
time is abandoned and hence there is no place for an extraneous, absolute
time variable that can be kept entirely separate from the arena where the
dynamical phenomena take place.) In its simplest version, this reformulation
amounts to replacing the configuration space $Q$ by the extended configuration
space $\, \mathbb{R} \times Q \,$ and the velocity phase space $TQ$ (the
tangent bundle of $Q$) by the extended velocity phase space $\, \mathbb{R}
\times TQ$, where $\mathbb{R}$ stands for the time axis. The usual momentum
phase space $T^* Q$ (the cotangent bundle of $Q$) admits two different
extensions: the simply extended phase space $\, \mathbb{R} \times T^* Q$,
where $\mathbb{R}$ represents the time variable, and the doubly extended
phase space $\, \mathbb{R} \times T^* Q \times \mathbb{R}$, where the first
copy of $\mathbb{R}$ represents the time variable whereas the second copy of
$\mathbb{R}$ represents an energy variable. This second extension is required
if one wants to maintain a symplectic structure, rather than just a contact
structure, for extended phase space, since energy is the physical quantity
canonically conjugate to time. A further generalization appears when one
considers mechanical systems in external gauge fields, since time-dependent
gauge transformations do not respect the direct product structure of the
extended configuration and phase spaces mentioned above. What does remain
invariant under such transformations are certain projections, namely the
projection from the extended configuration space onto the time axis, the
projections from the various extended phase spaces onto extended
configuration space and, finally, the projection from the doubly
extended to the simply extended phase space which amounts to
``forgetting the additional energy variable''.

In passing to field theory, we must replace the time axis $\mathbb{R}$
by the space-time manifold~$M$, the extended configuration space
$\, \mathbb{R} \times Q \,$ by the configuration bundle~$E$ over~$M$
introduced above and the extended velocity phase space $\, \mathbb{R}
\times TQ \,$ by the \emph{jet bundle}~$JE$ of~$E$.\footnote{We consider
only first order jet bundles and therefore omit the index ``1'' used by
many authors.} It is well known that $JE$ is~-- unlike the tangent
bundle of a manifold~-- in general only an affine bundle over~$E$
(of fiber dimension $Nn$) and not a vector bundle; the corresponding
difference vector bundle over~$E$ (also of fiber dimension $Nn$) will
\linebreak
be called the \emph{linearized jet bundle} of $E$ and be denoted by
$\vec{J} E$. This leads to the possibility of forming two kinds of dual:
the linear dual of~$\vec{J} E$, denoted here by $\vec{J}^\ast \! E$,
\linebreak
and the affine dual of~$JE$, denoted here by $J^\star \>\!\! E$; both of
them are vector bundles over~$E$ (of fiber dimension $Nn$ and $\, Nn+1$,
respectively). Even more important are their twisted versions, obtained
by taking the tensor product with the line bundle of volume forms on $M$,
pulled back to $E$ via $\pi$: this gives rise to the twisted linear dual
of~$\vec{J} E$, called \emph{ordinary multiphase space} and denoted here
by $\vec{J}\upoast E$, and the twisted affine dual of~$JE$, called
\emph{extended multiphase space} and denoted here by $J\upostar E$;
both of them, once again, are vector bundles over $E$ (of fiber dimension
$Nn$ and $\, Nn+1$, respectively). \linebreak The former replaces the simply
extended phase space $\, \mathbb{R} \times T^* Q \,$ of mechanics whereas
the latter replaces the doubly extended phase space $\, \mathbb{R} \times
T^* Q \times \mathbb{R} \,$ of mechanics. \linebreak Moreover, in both
cases (twisted or untwisted), there is a natural projection $\eta$ that,
as in mechanics, can be interpreted as ``forgetting the additional energy
variable'': it turns $J\upostar E$ into an affine line bundle over
$\vec{J}\upoast E$ and, similarly, $J^\star \>\!\! E$ into an affine line
bundle over $\vec{J}^\ast \! E$. The most remarkable property of extended
multiphase space is that it is an \emph{exact multisymplectic manifold}:
it carries a naturally defined \emph{multicanonical form} $\theta$, of
degree $n$, whose exterior derivative is the \emph{multisymplectic form}
$\omega$, of degree $n+1$, replacing the canonical form $\theta$ and the
symplectic form $\omega$, respectively, on the doubly extended phase
space $\, \mathbb{R} \times T^* Q \times \mathbb{R} \,$ of mechanics.

The global construction of the first order jet bundle $JE$ and the
linearized first order jet bundle $\vec{J} E$ associated with a given
fiber bundle $E$ over a manifold $M$, as well as that of the various
duals mentioned above, is quite easy to understand. (Higher order jet
bundles are somewhat harder to deal with, but we won't need them in
this paper.) Given a point $e$ in~$E$ with base point $\, x = \pi(e) \,$
in~$M$, the fiber $J_e^{} E$ of~$JE$ at~$e$ consists of all linear maps
from the tangent space $T_x^{} M$ of the base space~$M$ at~$x$ to the
tangent space $T_e^{} E$ of the total space~$E$ at~$e$ whose composition
with the tangent map $\, T_e^{} \pi: T_e^{} E  \rightarrow T_x^{} M \,$
to the projection $\, \pi: E \rightarrow M \,$ gives the identity
on~$T_x^{} M$:
\begin{equation} \label{eq:JETB1}
 J_e^{} E~=~\{ \, u_e^{} \smin L(T_x^{} M,T_e^{} E) \, / \,
                  T_e^{} \pi \smcirc u_e^{}
                  = \mathrm{id}_{T_x^{} M}^{} \, \}~.
\end{equation}
Thus the elements of $J_e^{} E$ are precisely the candidates for the
tangent maps at~$x$ to (local) sections $\varphi$ of the bundle~$E$
satisfying $\, \varphi(x) = e$. Obviously, $J_e E$ is an affine subspace
of the vector space $\, L(T_x^{} M, T_e^{} E) \,$ of all linear maps from
$T_x^{} M$ to the tangent space $T_e^{} E$, the corresponding difference
vector space being the vector space of all linear maps from~$T_x^{} M$ to
the vertical subspace~$V_e^{} E$:
\begin{equation} \label{eq:LJETB1}
 \vec{J}_e^{} E~=~L(T_x^{} M,V_e^{} E)~.
\end{equation}
The jet bundle $JE$ thus defined admits two different projections, namely
the \emph{target projection} $\, \tau_{JE}^{}: JE \rightarrow E \,$ and
the \emph{source projection} $\, \sigma_{JE}^{}: JE \rightarrow M \,$
which is simply its composition with the original projection $\pi$,
that~is, $\sigma_{JE}^{} = \pi \smcirc \tau_{JE}^{}$.
It is easily shown that $JE$ is a fiber bundle over~$M$ with respect
to~$\sigma_{JE}^{}$, in general without any additional structure,
but it is an affine bundle over~$E$ with respect to~$\tau_{JE}^{}$,
the corresponding difference vector bundle being the vector bundle
over~$E$ of linear maps from the pull-back of the tangent bundle of
the base space by the projection $\pi$ to the vertical bundle of $E$:
\begin{equation} \label{eq:LJETB2}
 \vec{J} E~=~L(\pi^* TM,VE)~.
\end{equation}
The affine structure of the jet bundle $JE$ over~$E$, as well as the linear
structure of the linearized jet bundle $\vec{J} E$ over~$E$, can also be
read off directly from local coordinate expressions. Namely, choosing local
coordinates $x^\mu$ for~$M$, local coordinates $q^i$ for~$Q$ and a local
trivialization of~$E$ induces naturally a local coordinate system
$(x^\mu,q^i,q_\mu^i)$ for $JE$, as well as a local coordinate system
$(x^\mu,q^i,\vec{q}_\mu^{\;i})$ for~$\vec{J} E$: such coordinates
will simply be referred to as \emph{adapted local coordinates}.
Moreover, a transformation to new local coordinates $x^{\prime\kappa}$
for~$M$, new local coordinates $q^{\prime k}$ for~$Q$ and a new local
trivialization of~$E$, according to
\begin{equation} \label{eq:CTJETB1}
 x^{\prime\kappa}~=~x^{\prime\kappa}(x^\mu)~~~,~~~
 q^{\prime k}~=~q^{\prime k}(x^\mu,q^i)
\end{equation}
induces naturally a transformation to new adapted local coordinates
$(x^{\prime\kappa},q^{\prime k},q_\kappa^{\prime k})$ for~$JE$ and
$(x^{\prime\kappa},q^{\prime k},\vec{q}_\kappa^{\;\prime k})$
for~$\vec{J} E$ given by eq.~(\ref{eq:CTJETB1}) and
\begin{equation} \label{eq:CTJETB2}
 q_\kappa^{\prime k}~=~q_\kappa^{\prime k}(x^\mu,q^i,q_\mu^i)~~~,~~~
 \vec{q}_\kappa^{\;\prime k}~
 =~\vec{q}_\kappa^{\;\prime k}(x^\mu,q^i,\vec{q}_\mu^{\;i})~,
\end{equation}
where
\begin{equation} \label{eq:CTJETB3}
 q_\kappa^{\prime k}~
 =~\frac{\partial x^\mu}{\partial x^{\prime\kappa}} \;
   \frac{\partial q^{\prime k}}{\partial q^i} \; q_\mu^i \; + \;
   \frac{\partial x^\mu}{\partial x^{\prime\kappa}} \;
   \frac{\partial q^{\prime k}}{\partial x^\mu}~~~,~~~
 \vec{q}_\kappa^{\;\prime k}~
 =~\frac{\partial x^\mu}{\partial x^{\prime\kappa}} \;
   \frac{\partial q^{\prime k}}{\partial q^i} \; \vec{q}_\mu^{\;i}~.
\end{equation}
Before going on, we pause to fix some notation concerning differential forms,
for which we shall in terms of local coordinates $x^\mu$ use the following
conventions:
\begin{equation} \label{eq:VOLDF1}
 d^{\,n} x~=~dx^1 \,\smwedge \ldots \smwedge\; dx^n~,
\end{equation}
\begin{equation} \label{eq:CMPDF1}
 d^{\,n} x_\mu^{}~=~i_{\partial_\mu}^{} \, d^{\,n} x~
 =~(-1)^{\mu-1} \, dx^1 \,\smwedge \ldots \smwedge\; dx^{\mu-1} \,\smwedge\;
                   dx^{\mu+1} \,\smwedge \ldots \smwedge\; dx^n~,
\end{equation}
\begin{equation} \label{eq:CMPDF2}
  d^{\,n} x_{\mu\nu}^{}~=~i_{\partial_\nu}^{} i_{\partial_\mu}^{} \, d^{\,n} x
  \quad \ldots \quad
  d^{\,n} x_{\mu_1 \ldots \mu_r}^{}~
  =~i_{\partial_{\mu_r}}^{} \ldots\, i_{\partial_{\mu_1}}^{} \, d^{\,n} x~.
\end{equation}
Then
\begin{equation} \label{eq:CMPDF3}
 i_{\partial_\mu}^{} \, d^{\,n} x_{\mu_1 \ldots\, \mu_r}^{}~
 =~d^{\,n} x_{\mu_1 \ldots\, \mu_r \mu}^{}~,
\end{equation}
whereas
\vspace{2mm}
\begin{equation} \label{eq:CMPDF4}
 dx^\kappa \,\smwedge\; d^{\,n} x_\mu^{}~=~\delta_\mu^\kappa \, d^{\,n} x~,
\vspace{2mm}
\end{equation}
\begin{equation} \label{eq:CMPDF5}
 dx^\kappa \,\smwedge\; d^{\,n} x_{\mu\nu}^{}~
 =~\delta_\nu^\kappa \, d^{\,n} x_\mu^{} \, - \,
   \delta_\mu^\kappa \, d^{\,n} x_\nu^{}~,
\end{equation}
\begin{equation} \label{eq:CMPDF6}
 dx^\kappa \,\smwedge\; d^{\,n} x_{\mu_1 \ldots\, \mu_r}^{}~
 =~\sum_{p=1}^r \, (-1)^{r-p} \, \delta_{\mu_p}^\kappa \,
   d^{\,n} x_{\mu_1 \ldots\, \mu_{p-1} \mu_{p+1} \ldots\, \mu_r}^{}~.
\end{equation}
Moreover, these (local) forms on $M$ are lifted to (local) forms on $E$ by
pull-back with the projection $\pi_E$, and later (local) forms on $E$ will
be lifted to (local) forms on total spaces of bundles over $E$ by pull-back
with the respective projection, without change of notation.

The dual $J_{}^\star \>\!\! E$ of the jet bundle $JE$ and the dual
$\vec{J}^\ast \! E$ of the linearized jet bundle $\vec{J} E$ are~%
obtained according to the standard rules for defining the dual of
an affine space and of a vector space, respectively. In particular,
these rules state that if $A$ is an affine space of dimension $k$
over $\mathbb{R}$, its dual $A^\star$ is the space $A(A,\mathbb{R})$
of affine maps from~$A$ to~$\mathbb{R}$, which is a vector space of
dimension $\, k+1$. Thus the dual or, more precisely, \emph{affine
dual} $J_{}^\star \>\!\! E$ of the jet bundle $JE$ and the dual or, more
precisely, \emph{linear dual} $\vec{J}^\ast \! E$ of~the linearized
jet bundle $\vec{J} E$ are obtained by defining their fiber over any
point $e$ in~$E$ to be the vector space
\begin{equation} \label{eq:DJETB01}
 J_e^\star E~=~\{ \, z_e^{}: J_e^{} E \longrightarrow \mathbb{R}~~
                     \mathrm{affine} \, \}~,
\end{equation}
and the vector space
\begin{equation} \label{eq:DLJETB01}
 \vec{J}_e^{\,\ast} E~
 =~\{ \, \vec{z}_e^{}: \vec{J}_e^{} E \longrightarrow \mathbb{R}~~
         \mathrm{linear} \, \}~,
\end{equation}
respectively. However, as mentioned before, the multiphase spaces of field
theory are defined with an additional twist, replacing the real line by
the one-dimensional space of volume forms on the base manifold $M$ at the
appropriate point. Thus the \emph{twisted (affine) dual} $J\upostar E$ of
the jet bundle $JE$ and the \emph{twisted (linear) dual} $\vec{J}\upoast E$
of the linearized jet bundle $\vec{J} E$ are obtained from the corresponding
ordinary (untwisted) duals by taking the tensor product with the line bundle
of volume forms on the base manifold $M$, pulled back to the total space $E$
via the projection $\pi$, i.e., we put
\begin{equation} \label{eq:CJETB01}
 J\upostar E~
 =~J_{}^\star \>\!\! E \otimes
   \pi^*(\bwedge^{\raisebox{-0.2ex}{${\scriptstyle n}$}} \, T^* M)~,
\end{equation}
and
\begin{equation} \label{eq:CLJETB01}
 \vec{J}\upoast E~
 =~\vec{J}^\ast E \otimes
   \pi^*(\bwedge^{\raisebox{-0.2ex}{${\scriptstyle n}$}} \, T^* M)~,
\vspace{1mm}
\end{equation}
respectively, which means that if $\, x = \pi(e)$, we set
\begin{equation} \label{eq:CJETB02}
 J_e\upostar E~
 =~\{ \, z_e^{}: J_e^{} E \longrightarrow
         \bwedge^{\raisebox{-0.2ex}{${\scriptstyle n}$}} \, T_x^* M~~
         \mathrm{affine} \, \}~,
\end{equation}
and
\begin{equation} \label{eq:CLJETB02}
 \vec{J}_e\upoast E~
 =~\{ \, \vec{z}_e^{}: \vec{J}_e^{} E \longrightarrow
         \bwedge^{\raisebox{-0.2ex}{${\scriptstyle n}$}} \, T_x^* M~~
         \mathrm{linear} \, \}~,
\end{equation}
respectively. As is the case for the jet bundle itself, the linearized jet
bundle and the various types of dual bundles introduced here all admit two
different projections, namely the \emph{target projection} $\tau_{\ldots}$
onto~$E$ and the \emph{source projection} $\sigma_{\ldots}$ onto~$M$ which
is simply its composition with the original projection $\pi$, that~is,
$\sigma_{\ldots} = \pi \smcirc \tau_{\ldots}$. It is easily shown that
all of them are fiber bundles over~$M$ with respect to~$\sigma_{\ldots}$,
in general without any additional structure, but~-- as stated before~--
they are vector bundles over~$E$ with respect to~$\tau_{\ldots}$.
The global linear structure of these bundles over~$E$ also becomes
clear in local coordinates. Namely, choosing local coordinates $x^\mu$
for~$M$, local coordinates $q^i$ for~$Q$ and a local trivialization
of~$E$ induces naturally not only local coordinate systems $(x^\mu,q^i,
q_\mu^i)$ for $JE$ and $(x^\mu,q^i,\vec{q}_\mu^{\;i})$ for $\vec{J} E$ 
but also local coordinate systems $(x_{\vphantom{i}}^\mu,q_{\vphantom{i}}^i,
p\;\!_i^\mu,p)$ both for $J_{}^\star \>\!\! E$ and for $J\upostar E$, as well
as local coordinate systems $(x_{\vphantom{i}}^\mu,q_{\vphantom{i}}^i,
p\>\!_i^\mu)$ both for $\vec{J}^\ast \! E$ and for $\vec{J}\upoast E$,
respectively: all these will again be referred to as \emph{adapted
local coordinates}. They are defined by requiring the dual pairing
between a point in~$J_{}^\star \>\!\! E$ or in~$J\upostar E$ with
coordinates $(x_{\vphantom{i}}^\mu,q_{\vphantom{i}}^i,p\>\!_i^\mu,p\;\!)$
and a point in~$JE$ with coordinates $(x^\mu,q^i,q_\mu^i)$ to be given by
\begin{equation} \label{eq:PJETB1}
 p\;\!_i^\mu q_\mu^i + \, p
\end{equation}
in the ordinary (untwisted) case and by
\begin{equation} \label{eq:PJETB2}
 \left( p\;\!_i^\mu q_\mu^i + \, p\;\! \right) \, d^{\,n} x
\end{equation}
in the twisted case, whereas the dual pairing between a point
in~$\vec{J}^\ast \! E$ or in~$\vec{J}\upoast E$ with coordinates
$(x_{\vphantom{i}}^\mu,q_{\vphantom{i}}^i,p\;\!_i^\mu)$ and a point
in~$\vec{J} E$ with coordinates $(x^\mu,q^i,\vec{q}_\mu^{\;i})$
should be given by
\begin{equation} \label{eq:PLJETB1}
 p\;\!_i^\mu \vec{q} _\mu^{\;i}
\end{equation}
in the ordinary (untwisted) case and by
\begin{equation} \label{eq:PLJETB2}
 p\;\!_i^\mu \vec{q} _\mu^{\;i} \; d^{\,n} x
\end{equation}
in the twisted case. Moreover, a transformation to new local coordinates
$x^{\prime\kappa}$ for~$M$, new local coordinates $q^{\prime k}$ for~$Q$
and a new local trivialization of~$E$, according to eq.~(\ref{eq:CTJETB1}),
induces naturally not only a transformation to new adapted local coordinates
$(x^{\prime\kappa},q^{\prime k},q_\kappa^{\prime k})$ for $JE$ and
$(x^{\prime\kappa},q^{\prime k},\vec{q} _\kappa^{\;\prime k})$
for~$\vec{J} E$, as given by eqs~(\ref{eq:CTJETB2}) and~(\ref{eq:CTJETB3}),
but also a transformation to new adapted local coordinates $(x^{\prime\kappa},
q^{\prime k},p\:\!_k^{\prime\kappa},p\:\!^\prime)$ both for $J_{}^\star \>\!\!
E$ and for $J\upostar E$, as well as a transformation to new adapted local
coordinates $(x^{\prime\kappa},q^{\prime k},p\;\!_k^{\prime\kappa})$ both
for~$\vec{J}^\ast \! E$ and  for~$\vec{J}\upoast E$, respectively:
they are given by
\begin{equation} \label{eq:CTDCJETB1}
 p\;\!_k^{\prime\kappa}~
 =~p\;\!_k^{\prime\kappa}(x_{\vphantom{i}}^\mu,
                          q_{\vphantom{i}}^i,p\;\!_i^\mu,p)~~~,~~~
 p\;\!^\prime~=~p\;\!^\prime (x_{\vphantom{i}}^\mu,
                              q_{\vphantom{i}}^i,p\;\!_i^\mu,p)~,
\end{equation}
where
\begin{equation} \label{eq:CTDJETB2}
 p\;\!_k^{\prime\kappa}~
 =~\frac{\partial x^{\prime\kappa}}{\partial x^\mu} \;
   \frac{\partial q^i}{\partial q^{\prime k}} \; p\;\!_i^\mu~~~,~~~
 p\;\!^\prime~
 =~p \; - \, \frac{\partial q^{\prime k}}{\partial x^\mu} \;
             \frac{\partial q^i}{\partial q^{\prime k}} \; p\;\!_i^\mu
\end{equation}
in the ordinary (untwisted) case and
\begin{equation} \label{eq:CTCJETB2}
 p\;\!_k^{\prime\kappa}~
 =~\det \Bigl( \frac{\partial x}{\partial x^\prime} \Bigr) \;
   \frac{\partial x^{\prime\kappa}}{\partial x^\mu} \;
   \frac{\partial q^i}{\partial q^{\prime k}} \; p\;\!_i^\mu~~~,~~~
 p\;\!^\prime~
 =~\det \Bigl( \frac{\partial x}{\partial x^\prime} \Bigr)
   \left( p \; - \, \frac{\partial q^{\prime k}}{\partial x^\mu} \;
                    \frac{\partial q^i}{\partial q^{\prime k}} \; p\;\!_i^\mu
          \right)
\end{equation}
in the twisted case. Finally, it is worth noting that the affine duals
$J_{}^\star \>\!\! E$ and $J\upostar E$ of~$JE$ contain line subbundles
$J_0^\star \>\!\! E$ and $J_0\upostar E$ whose fiber over any point $e$
in~$E$ consists of the constant (rather than affine) maps from $J_e^{} E$
to $\mathbb{R}$ and to $\bwedge^{\raisebox{-0.2ex}{${\scriptstyle n}$}} \,
T_x^* M$, respectively, and the corresponding quotient vector bundles
over~$E$ can be naturally identified with the respective duals
$\vec{J}^\ast \! E$ and $\vec{J}\upoast E$ of~$\vec{J} E$, i.e.,
we have
\begin{equation} \label{eq:DJETB03}
 J_{}^\star \>\!\! E / J_0^\star \>\!\! E~\cong~\vec{J}^\ast \! E~\cong~
 L(VE,\pi^* TM)~,
\end{equation}
and
\begin{equation} \label{eq:CJETB03}
 J\upostar E / J_0\upostar E~\cong~\vec{J}\upoast E~\cong~
 L(VE,\pi^*(\bwedge^{\raisebox{-0.2ex}{${\scriptstyle n-1}$}} \, T^* M))~,
\end{equation}
respectively. This shows that, in both cases, the corresponding projection
onto the quotient amounts to ``forgetting the additional energy variable''
since it takes a point with coordinates $(x_{\vphantom{i}}^\mu,%
q_{\vphantom{i}}^i,p\;\!_i^\mu,p)$ to the point with coordinates
$(x_{\vphantom{i}}^\mu,q_{\vphantom{i}}^i,p\;\!_i^\mu)$; it will be
denoted by $\eta$ (as a reminder for the fact that it projects the
extended multiphase space to the ordinary one) and is easily seen
to turn $J_{}^\star \>\!\! E$ and $J\upostar E$ into affine line bundles
over $\vec{J}^\ast \! E$ and over $\vec{J}\upoast E$, respectively.
  
An alternative but equivalent description of the extended multiphase
space of field theory is as a certain bundle of differential forms on
the total space~$E$ of the configuration bundle, namely the bundle
${\bwedge\,}_{n-1}^{\raisebox{-0.2ex}{${\scriptstyle n}$}} \, T^* E$ of
$(n\!-\!1)$-horizontal $n$-forms on $E$, that is, of $n$-forms on $E$
that vanish whenever one inserts at least two vertical vectors. In fact,
there is a canonical isomorphism
\begin{equation} \label{eq:CJETB04}
 \Phi:~{\bwedge\,}_{n-1}^{\raisebox{-0.2ex}{${\scriptstyle n}$}} \, T^* E~
 \stackrel{\cong}{\longrightarrow}~J\upostar E
\end{equation}
of vector bundles over $E$ that can be defined explicitly as follows:
given any point $e$ in~$E$ with base point $\, x = \pi(e) \,$ in~$M$
and any $(n\!-\!1)$-horizontal $n$-form $\, \alpha_e^{} \smin\,
{\bwedge\,}_{n-1}^{\raisebox{-0.2ex}{${\scriptstyle n}$}} \, T_e^* E$,
together with a jet $\, u_e^{} \smin J_e^{} E$, we can use $u_e$, which
is a linear map from $T_x^{} M$ to $T_e^{} E$, to pull back the $n$-form
$\alpha_e^{}$ on $T_e^{} E$ to an $n$-form $u_e^* \alpha_e^{}$ on $T_x^{} M$.
Obviously, $u_e^* \alpha_e^{}$ is an affine function of $u_e^{}$ as $u_e^{}$
varies over the affine space $J_e^{} E$ because it is actually a linear
function of $u_e^{}$ when $u_e^{}$ is allowed to vary over the entire
vector space $\, L(T_x^{} M,T_e^{} E)$ \linebreak (the restriction of
a linear map between two vector spaces to an affine subspace of its
domain is an affine map). Thus putting
\begin{equation} \label{eq:CJETB05}
 \Phi_e^{}(\alpha_e^{}) \cdot u_e^{}~=~u_e^* \alpha_e^{}
\end{equation}
defines a map $\; \Phi_e^{}: {\bwedge\,}_{n-1}^{\raisebox{-0.2ex}%
{${\scriptstyle n}$}} \, T_e^* E \rightarrow J_e^\star E \;$ which
is evidently linear and, as $e$ varies over~$E$, provides the desired
isomorphism (\ref{eq:CJETB04}). Further details can be found in
Ref.~\cite{GIM}. The~importance of this canonical isomorphism is
due to the fact that it provides a natural way to introduce a
\emph{multicanonical form} $\theta$ and a \emph{multisymplectic
form} $\omega$ on extended multiphase space which play a similar
role in field theory as the canonical form $\theta$ and the symplectic
two-form $\omega$ on cotangent bundles in mechanics. Namely, $\theta$ is
an $n$-form that can be defined intrinsically by using the tangent map
$\, T \tau_{J\smupostar E}^{}: T(J\upostar E) \rightarrow TE \,$ to the
bundle projection $\, \tau_{J\smupostar E}^{} : J\upostar E \rightarrow E$,
as follows. Given a point $\, z \smin J\upostar E \,$ with base point
$\, e = \tau_{J\smupostar E}^{}(z) \,$ in $E$ and $n$ tangent vectors
$\, w_1\,,\ldots, w_n \,$ to $J\upostar E$ at $z$, put
\begin{equation} \label{eq:MCANF01}
 \theta_z^{}(w_1,\ldots,w_n)~
 =~(\Phi_e^{-1}(z))(T_z^{} \tau_{J\smupostar E}^{} \cdot w_1 \,,\, \ldots \,,\,
                    T_z^{} \tau_{J\smupostar E}^{} \cdot w_n)~.
\end{equation}
Moreover, $\omega$ is an $(n+1)$-form which, as in mechanics, is defined
to be the negative of the exterior derivative of $\theta$:
\begin{equation} \label{eq:MSYMF01}
 \omega~=~- \, d\theta~.
\end{equation}
Another important object that can be defined globally both on extended and
ordinary multiphase space is the \emph{scaling} or \emph{Euler vector field}
which we shall denote here by $\Sigma$. Its definition is based exclusively
on the fact that $J\upostar E$ and $\vec{J}\upoast E$ are total spaces of
vector bundles over $E$. In fact, given any vector bundle $V$ over $E$,
$\Sigma_V^{}$ (which we shall simply denote by $\Sigma$ when there is
no danger of confusion) is defined to be the fundamental vector field
associated with the action of $\mathbb{R}$, considered as a commutative
group under addition, by scaling transformations on the fibers:
\[
 \begin{array}{ccc}
  \mathbb{R} \times V & \longrightarrow &         V          \\[2mm]
      (\lambda,v)     &   \longmapsto   & \exp(\lambda) \, v
 \end{array}~.
\]
Thus $\Sigma$ is simply that vertical vector field on $V$ which, under
identification of the vertical tangent spaces to $V$ with the fibers of $V$
itself typical for vector bundles, becomes the identity on $V$:
\[
 \Sigma(v)~=~\frac{d}{d\lambda} \; \exp(\lambda) \, v  \, \bigg|_{\lambda=0}~
           =~v~.
\]
In adapted local coordinates, the isomorphism $\Phi$ can be defined by the
requirement that the $(n\!-\!1)$-horizontal $n$-form on $E$ corresponding
to the point in~$J\upostar E$ with coordinates $(x_{\vphantom{i}}^\mu,%
q_{\vphantom{i}}^i,p\;\!_i^\mu,p)$ is explicitly given by
\begin{equation} \label{eq:CJETB10}
 p\;\!_i^\mu \; dq^i \,\smwedge\; d^{\,n} x_\mu^{} \, + \; p \; d^{\,n} x~.
\end{equation}
The tautological nature of the definition of $\theta$ then becomes apparent
by realizing that exactly the same expression represents the multicanonical
form $\theta$:
\begin{equation} \label{eq:MCANF02}
 \theta~=~p\;\!_i^\mu \; dq^i \,\smwedge\; d^{\,n} x_\mu^{} \, + \; 
          p \; d^{\,n} x~.
\end{equation}
Taking the exterior derivative yields
\begin{equation} \label{eq:MSYMF02}
 \omega~=~dq^i \,\smwedge\; dp\;\!_i^\mu \,\smwedge\; d^{\,n} x_\mu^{} \, - \;
          dp \,\:\smwedge\; d^{\,n} x~.
\end{equation}
Moreover, the scaling vector fields on $J\upostar E$ and on $\vec{J}\upoast E$
are given by
\begin{equation} \label{eq:SVFEMPS}
 \Sigma~=~p\;\!_i^\mu \, \frac{\partial}{\partial p\;\!_i^\mu} \, + \,
          p \: \frac{\partial}{\partial p}
\end{equation}
and by
\begin{equation} \label{eq:SVFOMPS}
 \Sigma~=~p\;\!_i^\mu \, \frac{\partial}{\partial p\;\!_i^\mu}
\vspace{2mm}
\end{equation}
respectively. Finally, we note the following relations, which will be used
later.
\begin{prp} \label{prp:SVFMCSF}
 The multicanonical form $\theta$, the multisymplectic form $\omega$ and the
 \mbox{scaling} or Euler vector field $\,\Sigma$ on extended multiphase space
 $J\upostar E$ satisfy the following relations:
 \begin{eqnarray}
  &L_\Sigma^{} \theta~=~\theta~.&                    \label{eq:MCANF03} \\[1mm]
  &L_\Sigma^{} \omega~=~\omega~.&                    \label{eq:MSYMF03} \\[1mm]
  &i_\Sigma^{} \theta~=~0~.&                         \label{eq:MCANF04} \\[1mm]
  &i_\Sigma^{} \omega~=~- \, \theta~.&               \label{eq:MSYMF04}
 \end{eqnarray}
\end{prp}
\textbf{Proof}~~Let $(\varphi_\lambda)_{\lambda \ssmin \mathbb{R}}$ denote
the one-parameter group of scaling transformations on $J\upostar E$ given
by $\, \varphi_\lambda(z) = \mathrm{e}^\lambda z$. Then by the formula
relating the Lie derivative of a differential form along a vector field
to the derivative of its pull-back under the flow of that vector field
(see, e.g.,~\cite[p.~91]{AM}) and the definition of $\theta$, we have
\begin{eqnarray*}
\lefteqn{\left( L_\Sigma^{} \theta \right)_z (w_1^{},\ldots,w_n^{})~
         =~\frac{\partial}{\partial \lambda}
          \left( \varphi_\lambda^\ast \theta \right)_z^{}
          (w_1^{},\ldots,w_n^{}) \, \bigg|_{\lambda=0}}         \hspace{5mm} \\
 &=&\!\! \frac{\partial}{\partial \lambda} \;
         \theta_{\varphi_\lambda(z)}
         \left( T_z^{} \varphi_\lambda^{} \cdot w_1^{} \, , \ldots ,
                T_z^{} \varphi_\lambda^{} \cdot w_n^{} \right)
         \bigg|_{\lambda=0}                                                  \\
 &=&\!\! \frac{\partial}{\partial \lambda} \;
         \Phi_e^{-1}(\varphi_\lambda^{}(z))
         \left( T_{\varphi_\lambda(z)}^{} \tau_{J\smupostar E}^{\phantom{j}}
                \cdot (T_z^{} \varphi_\lambda^{} \cdot w_1^{}) \, , \ldots ,
                T_{\varphi_\lambda(z)}^{} \tau_{J\smupostar E}^{\phantom{j}}
                \cdot (T_z^{} \varphi_\lambda^{} \cdot w_n^{}) \right)
         \bigg|_{\lambda=0}                                                  \\
 &=&\!\! \frac{\partial}{\partial \lambda} \;
         \Phi_e^{-1}(\mathrm{e}^\lambda z)
         \left( T_z^{} ( \tau_{J\smupostar E}^{\phantom{j}} \smcirc
                         \varphi_\lambda^{} ) \cdot w_1^{} \, , \ldots ,
                T_z^{} ( \tau_{J\smupostar E}^{\phantom{j}} \smcirc
                         \varphi_\lambda^{} ) \cdot w_n^{} \right)
         \bigg|_{\lambda=0}                                                  \\
 &=&\!\! \frac{\partial}{\partial \lambda} \;
         \mathrm{e}^\lambda \, \Phi_e^{-1}(z)
         \left( T_z^{} \tau_{J\smupostar E}^{\phantom{j}} \cdot w_1^{} \, ,
                \ldots ,
                T_z^{} \tau_{J\smupostar E}^{\phantom{j}} \cdot w_n^{} \right)
         \bigg|_{\lambda=0}                                                  \\
 &=&\!\! \frac{\partial}{\partial \lambda} \;
         \mathrm{e}^\lambda \,
         \theta_z^{}(w_1^{},\ldots,w_n^{}) \, \bigg|_{\lambda=0}        \\[2mm]
 &=&\!\! \theta_z^{}(w_1^{},\ldots,w_n^{})~,
\end{eqnarray*}
which proves eq.~(\ref{eq:MCANF03}) and also eq.~(\ref{eq:MSYMF03}) since
$L_\Sigma^{}$ commutes with the exterior derivative. Next, observe that
with respect to the target projection of $J\upostar E$ onto $E$, $\Sigma$ is
vertical whereas $\theta$ is horizontal, which imples eq.~(\ref{eq:MCANF04}).
Combining these two equations, we finally get
\[
 \theta~=~L_\Sigma^{} \theta~
 =~d \left( i_\Sigma \theta \right) \, + \, i_\Sigma \, d \>\! \theta~
 =~- \, i_\Sigma \omega~,
\]
proving eq.~(\ref{eq:MSYMF04}).
\PCPqed

We note here that the existence of the canonically defined forms $\theta$
and $\omega$ is what distinguishes the twisted affine dual $J\upostar E$
from the ordinary affine dual $J_{}^\star \>\!\! E$ of $JE$.

Using the jet bundle $JE$ and the multiphase spaces $\vec{J}\upoast E$ and
$J\upostar E$ associated with a given fiber bundle $E$ over space-time $M$,
one can develop a general covariant Lagrangian and Hamiltonian formalism
for field theories whose configurations are sections of $E$. \linebreak
For example, the Lagrangian function of mechanics is replaced by a
\emph{Lagrangian density} $\mathcal{L}$, which is a function on $JE$
with values in the volume forms on space-time, so that one can integrate
it to compute the action functional and formulate a variational principle.
It gives rise to a \emph{covariant Legendre transformation} which replaces
that of mechanics and comes in two variants, both defined by an appropriate
notion of vertical derivative or fiber derivative: one of them is a fiber
preserving smooth map $\; \vec{\mathbb{F}} \mathcal{L} : JE \rightarrow
\vec{J}\upoast E$ \linebreak and the other a fiber preserving smooth map
$\; \mathbb{F} \mathcal{L} : JE \rightarrow J\upostar E$; of course, the
former is obtained from the latter by composition with the natural projection
$\eta$ from~$J\upostar E$ onto~$\vec{J}\upoast E$ mentioned above. When
$\vec{\mathbb{F}} \mathcal{L}$ is a local/\,global diffeomorphism, the
Lagrangian $\mathcal{L}$ is called \emph{regular}/\,\emph{hyperregular}.
On the other hand, the Hamiltonian function of mechanics is replaced by a
\emph{Hamiltonian density} $\mathcal{H}$, which is a section of extended
\mbox{multi}\-phase space $J\upostar E$ as an affine line bundle over
ordinary multiphase space $\vec{J}\upoast E$. \linebreak Once again,
any such section gives rise to a \emph{covariant Legendre transformation},
defined by an appropriate notion of vertical derivative or fiber derivative:
it is a fiber preserving smooth map $\; \mathbb{F} \mathcal{H} :
\vec{J}\upoast E \rightarrow JE$. When $\mathbb{F} \mathcal{H}$ is a
local/\,global diffeomorphism, the Hamiltonian $\mathcal{H}$ is called
\emph{regular}/\,\emph{hyperregular}. In any case, pulling back $\theta$
and $\omega$ from $J\upostar E$ to $JE$ via $\mathbb{F} \mathcal{L}$
generates the \emph{Poincar\'e-Cartan forms} $\theta_{\mathcal{L}}$
and $\omega_{\mathcal{L}}$ on $JE$, and similarly, pulling them back
from $J\upostar E$ to $\vec{J}\upoast E$ via $\mathcal{H}$ generates the
forms $\theta_{\mathcal{H}}$ and $\omega_{\mathcal{H}}$ on $\vec{J}\upoast E$.
\linebreak
As in mechanics, the Lagrangian and Hamiltonian formulations turn out to
be completely equivalent in the hyperregular case, with $\vec{\mathbb{F}}
\mathcal{L}$ and $\mathbb{F} \mathcal{H}$ being each other's inverse.
For more details on these and related matters, the reader may consult
Ref.~\cite{CCI} and, in particular, Ref.~\cite{GIM} -- except for the
direct construction of the Legendre transformation $\vec{\mathbb{F}}
\mathcal{H}$ associated with a Hamiltonian $\mathcal{H}$, which was
first derived in Ref.~\cite{Ro}; see also Ref.~\cite{FR2}. There is
also a generalization of the Hamilton-Jacobi equation to the field
theoretical situation; the reader may consult the extensive review
by Kastrup \cite{Kas} as a starting point for this direction.

\section{Poisson forms and their Poisson brackets}

The constructions exposed in the previous section have identified the
extended multiphase space of field theory as an example of a multisymplectic
manifold.
\begin{dfn} \label{dfn:MSPLMAN}
 A \textbf{multisymplectic manifold\/} is a manifold $P$ equipped
 with a non-degenerate closed $(n+1)$-form $\omega$, called the
 \textbf{multisymplectic form}.
\end{dfn}
\textbf{Remark}~~This definition is deliberately vague as to the meaning
of the term ``non-degenerate'', at least when $\, n > 1$. The standard
interpretation is that the kernel of $\omega$ on vectors should vanish,
that is,
\begin{equation} \label{eq:NONDEG1}
 i_X^{} \omega~=~0~~\Longrightarrow~~X~=~0 \qquad
 \mbox{for vector fields $X$}~.
\end{equation}
Note that, of course, no such conclusion holds for multivector fields, that
is, the kernel of $\omega$ on multivectors is non-trivial. (This is true
even for symplectic forms which vanish on certain bivectors, for example
on those that represent two-dimensional isotropic subspaces.) However,
the condition (\ref{eq:NONDEG1}) alone is too weak and it is not clear
what additional algebraic constraints should be imposed on $\omega$.
A first attempt in this direction has been made by Martin \cite{Ma1,Ma2},
but his conditions are too restrictive and do not seem to agree with what
is needed in applications to field theory. More recently, a promising
proposal has been made by Cantrijn, Ibort and de Le\'on \cite{CIL}
which seems to come close to a convincing definition of the concept
of a multisymplectic manifold. Fortunately, there is no need to enter
this discussion here since the ``minimal'' requirement of non-degeneracy
formulated in eq.\ (\ref{eq:NONDEG1}) is sufficient for our purposes and
will be used here to provide a working definition.

In what follows, we shall make extensive use of the basic operations of
calculus on manifolds involving multivector fields and differential forms,
namely the Schouten bracket between multivector fields, the contraction
of differential forms with multivector fields and the Lie derivative of
differential forms along multivector fields. For the convenience of the
reader, the relevant formulae are summarized in Appendix A; in particular,
eqs~(\ref{eq:LDFMVF}) and~(\ref{eq:LXiY-iYLX}) will be used constantly and
often without further mention.

On multisymplectic manifolds, there are special classes of multivector
fields and of differential forms:
\begin{dfn} \label{dfn:HAMMVFF}
 An $r$-multivector field $X$ on a multisymplectic manifold $P$ is called
 \textbf{locally Hamiltonian\/} if $\; i_X \omega \,$ is closed, or
 equivalently, if
 \begin{equation} \label{eq:LHMVF}
  L_X^{} \omega~=~0~,
 \end{equation}
 and it is called \textbf{globally Hamiltonian\/} or simply
 \textbf{Hamiltonian\/} if $\; i_X \omega \,$ is exact, i.e.,
 if there exists an $(n-r)$-form $f$ on $P$ such that
 \begin{equation} \label{eq:GHMVF}
  i_X^{} \omega~=~df~.
 \end{equation}
 In this case, we say that $f$ is \textbf{associated\/} with $X$ or
 \textbf{corresponds\/} to $X$. \\[1mm]
 Conversely, an $(n-r)$-form $f$ on a multisymplectic manifold $P$ is
 called \textbf{Hamiltonian\/} if there exists an $r$-multivector field
 $X$ on $P$ such that
 \begin{equation} \label{eq:HFORM}
  i_X^{} \omega~=~df~.
 \end{equation}
 In this case, we say that $X$ is \textbf{associated\/} with $f$ or
 \textbf{corresponds\/} to $f$.
\end{dfn}
\textbf{Remark}~~As mentioned before, the kernel of $\omega$ on multivectors
is non-trivial, so the correspondence between Hamiltonian multivector fields
and Hamiltonian forms is not unique (in either direction). Moreover, by far
not every form is Hamiltonian. \linebreak In particular, as first shown in
special examples by Kijowski \cite{Kij} and then more systematically by
Kanatchikov \cite{Ka1}, although in a somewhat different context, there
are restrictions on the allowed multimomentum dependence of the coefficient
functions. Of course, every closed form is Hamiltonian (the corresponding
Hamiltonian multivector field vanishes identically). Below we will give
more interesting examples to show that the definition is not empty.
\begin{prp} \label{prp:HAMMVFF}
 The Schouten bracket of any two locally Hamiltonian multivector fields
 $X$ and $Y$ on a multisymplectic manifold $P$ is a globally Hamiltonian
 multivector field $\, [X,Y]$ on $P$ whose associated Hamiltonian form can,
 up to sign, be chosen to be the double contraction $\, i_X^{} i_Y^{} \omega$.
 More precisely, assuming $X$ to be of degree $r$ and $Y$ to be of
 degree $s$, we have
 \begin{equation} \label{eq:SCHBRPB1}
  i_{[X,Y]}^{} \>\! \omega~
  =~(-1)^{(r-1)s} \, d \left( i_X^{} i_Y^{} \omega \right)\!~.
 \end{equation}
 In particular, this implies that under the Schouten bracket, the space
 $\mathfrak{X}_{LH}^\wedge(P)$ of locally Hamiltonian multivector fields on
 $P$ is a subalgebra of the Lie superalgebra $\mathfrak{X}^\wedge(P)$ of all
 multivector fields on $P$, containing the space $\mathfrak{X}_H^\wedge(P)$
 of globally Hamiltonian multivector fields, as well as the (smaller) space
 $\mathfrak{X}_0^\wedge(P)$ of multivector fields taking values in the
 kernel of $\,\omega$, as ideals:  if $X$ is locally Hamiltonian, then
 \begin{equation} \label{eq:KERNI}
  i_\xi^{} \>\! \omega~=~0 \quad \Longrightarrow \quad
  i_{[\xi,X]}^{} \>\! \omega~=~0~.
 \end{equation}
\end{prp}
\textbf{Proof}~~According to eqs (\ref{eq:LXiY-iYLX}) and (\ref{eq:LDFMVF}),
we have for any two multivector fields $X$ of degree $r$ and $Y$ of degree $s$,
\begin{eqnarray*}
 i_{[X,Y]} \, \omega \!\!
 &=&\!\! (-1)^{(r-1)s} \, L_X^{} i_Y^{} \omega \, - \, i_Y^{} L_X^{} \omega  \\
 &=&\!\! (-1)^{(r-1)s} \, d \left( i_X^{} i_Y^{} \omega \right) \, + \,
         (-1)^{(r-1)(s-1)} \, i_X^{} \, d \left( i_Y^{} \omega \right) \, - \,
         i_Y^{} L_X^{} \omega                                                \\
 &=&\!\! (-1)^{(r-1)s} \, d \left( i_X^{} i_Y^{} \omega \right) \, + \,
         (-1)^{(r-1)(s-1)} \, i_X^{} L_Y^{} \omega \, - \,
         i_Y^{} L_X^{} \omega~,
\end{eqnarray*}
since $\, d \>\! \omega = 0$, showing that if $X$ and $Y$ are both locally
Hamiltonian, then $[X,Y]$ is globally Hamiltonian and eq.\ (\ref{eq:SCHBRPB1})
holds. \PCPqed
\begin{dfn} \label{dfn:POISF}
 A Hamiltonian form $f$ on a multisymplectic manifold $P$ is called a
 \textbf{Poisson form\/} if its contraction with any multivector field
 $\xi$ on $P$ taking values in the kernel of $\,\omega$ vanishes, i.e.,
 if
 \begin{equation} \label{eq:KERN1}
  i_\xi^{} \, \omega~=~0~~\Longrightarrow~~i_\xi^{} f~=~0~,
 \end{equation}
 and is called a \textbf{weak Poisson form\/} it its contraction with any
 multivector field $\xi$ on $P$ taking values in the kernel of $\,\omega$ 
 is a closed form, i.e., if
 \begin{equation} \label{eq:KERN2}
  i_\xi^{} \, \omega~=~0~~\Longrightarrow~~d \left( i_\xi^{} f \right) \; =~0~.
 \end{equation}
\end{dfn}
\begin{dfn} \label{dfn:EXMSM}
 An \textbf{exact multisymplectic manifold\/} is a multisymplectic manifold
 whose multisymplectic form $\omega$ is the exterior derivative of a Poisson
 form:
 \begin{equation} \label{eq:EXMSM}
  \omega~=~- \, d \>\! \theta~.
 \end{equation}
 \begin{equation} \label{eq:KERN3}
  i_\xi^{} \, \omega~=~0~~\Longrightarrow~~i_\xi^{} \, \theta~=~0~.
 \end{equation}
 We shall call $\theta$ the \textbf{multicanonical form}.
\end{dfn}
\textbf{Remark}~~It is an immediate consequence of Proposition~\ref%
{prp:SVFMCSF}, in particular of eq.~(\ref{eq:MSYMF04}), that the extended
multiphase space of field theory is an exact multisymplectic manifold.
However, the condition that the kernel of $\theta$ should contain that
of~$\omega$ is non-trivial in the sense that it is not always possible
to modify a potential of an exact form by adding an appropriate closed form
so as to achieve the desired inclusion of the kernels, as the following
counterexample will show.\footnote{This example is due to M.~Bordemann.}
Consider the three-sphere $S^3$ as the total space of the Hopf bundle,
a principal $U(1)$-bundle over the two-sphere $S^2$, and let $\xi$ be
the fundamental vector field of the $U(1)$ group action on $S^3$ and
$\alpha$ be the canonical connection $1$-form on $S^3$. Then $i_\xi^{}
\alpha = 1$ and $i_\xi d\alpha = 0$. We want to modify $\alpha$
by some closed form $\beta$ so that $i_\xi (\alpha+\beta) = 0$.
But $S^3$ is simply connected, so $d\beta=0$ implies that there is a
function $f$ with $df=\beta$. Hence we are looking for a function $f$
on $S^3$ that satisfies $i_\xi df = -1$. But $S^3$ is compact, so $f$
must have at least two critical points (a maximum and a minimum), and
we arrive at a contradiction. In other words, we cannot modify the
potential $\alpha$ of $d \alpha$ in such a way that the kernel of
$d\alpha$ is contained in the kernel of the modified potential.
\begin{dfn} \label{dfn:POISB}
 Let $P$ be an exact multisymplectic manifold. Given any two (weak)
 Poisson forms $f$ of degree $n-r$ and $g$ of degree $n-s$ on $P$,
 their \textbf{Poisson bracket\/} is defined to be the $(n+1-r-s)$-form
 on $P$ given by
 \begin{equation} \label{eq:POISB1}
  \{f,g\}~=~- \, L_X^{} g \, + \, (-1)^{(r-1)(s-1)} L_Y^{} f \,
            - \, (-1)^{(r-1)s} L_{X \wedge\>\! Y} \, \theta~,
 \end{equation}
 or equivalently,
 \begin{equation} \label{eq:POISB2}
  \begin{array}{rcl}
   \{f,g\} \!&=&\! (-1)^{r(s-1)} \, i_Y^{} i_X^{} \omega \vphantom{\Bigl(} \\
             & & \mbox{} + \; d \,
                 \Bigl( (-1)^{(r-1)(s-1)} \, i_Y^{} f \, - \, i_X^{} g \, - \,
                        (-1)^{(r-1)s} \, i_Y^{} i_X^{} \theta \, \Bigr)~,
  \end{array}
 \end{equation}
 where $X$ and $\,Y$ are Hamiltonian multivector fields associated with $f$
 and with $g$, \linebreak respectively.
\end{dfn}
\textbf{Remark}~~This Poisson bracket is an extension of the one between
Hamiltonian $(n-1)$-forms introduced by two of the present authors in an
earlier article \cite{FR1}, except for the fact that when $f$ and $g$ are
$(n-1)$-forms, $X$ and $Y$ are vector fields and are uniquely determined by
$f$ and $g$, so there is no need to impose restrictions on the contraction of
$f$ and $g$ with multivector fields taking values in the kernel of $\omega$:
the definition given in Ref.\ \cite{FR1} works for all Hamiltonian $(n-1)$-%
forms and not just for (weak) Poisson $(n-1)$-forms.
\begin{prp}
 The Poisson bracket introduced above is well defined and closes, i.e.,
 when $f$ and $g$ are Poisson forms, $\{f,g\}$ does not depend on the
 choice of the Hamiltonian multivector fields $X$ and $Y$ used in its
 definition and is itself a Poisson form.  The same statement holds if
 the term ``Poisson form'' is replaced by the term ``weak Poisson form''.
 Moreover, we have the important relation
 \begin{equation} \label{eq:POISB3}
  i_{[Y,X]}^{} \omega~=~d \, \{f,g\}~,
 \end{equation}
 i.e., if  $X$ is a Hamiltonian multivector field associated with $f$ and
 $Y$ is a Hamiltonian multivector field associated with $g$, then $[Y,X]$
 is a Hamiltonian multivector field associated with $\{f,g\}$.
\end{prp}
\textbf{Proof}~~We begin by using eq.~(\ref{eq:LDFMVF}) to show that, for
any two Hamiltonian forms $f$ of degree $n-r$ and $g$ of degree $n-s$ with
associated Hamiltonian multivector fields $X$ and $Y$, respectively, the
expressions on the rhs of eqs~(\ref{eq:POISB1}) and~(\ref{eq:POISB2})
coincide:
\begin{eqnarray*}
\lefteqn{- \, L_X^{} g \, + \, (-1)^{(r-1)(s-1)} L_Y^{} f \,
         - \, (-1)^{(r-1)s} L_{X \wedge\>\! Y} \, \theta}  \hspace{1cm} \\[2mm]
 &=&\!\! \mbox{} - \, d \left( i_X^{} g \right)
                 + \, (-1)^r \, i_X^{} dg                               \\[1mm]
 & &\!\! \mbox{} + \, (-1)^{(r-1)(s-1)} \, d \left( i_Y^{} f \right)
                 - \, (-1)^{(r-1)(s-1)+s} \, i_Y^{} df                  \\[1mm]
 & &\!\! \mbox{} - \, (-1)^{(r-1)s} \,
                      d \left( i_{X \wedge Y}^{} \, \theta \right)
                 - \, (-1)^{r(s-1)} \, i_{X \wedge\>\! Y}^{} \, \omega  \\[2mm]
 &=&\!\! \mbox{} - \, d \left( i_X^{} g \right)
                 + \, (-1)^{rs+r} \, i_Y^{} i_X^{} \omega               \\[1mm]
 & &\!\! \mbox{} + \, (-1)^{(r-1)(s-1)} \, d \left( i_Y^{} f \right)
                 + \, (-1)^{rs-r} \, i_Y^{} i_X^{} \omega               \\[1mm]
 & &\!\! \mbox{} - \, (-1)^{(r-1)s} \,
                      d \left( i_Y^{} i_X^{} \, \theta \right)
                 - \, (-1)^{rs-r} \, i_Y^{} i_X^{} \omega               \\[2mm]
 &=&\!   (-1)^{r(s-1)} \, i_Y^{} i_X^{} \omega \vphantom{\Bigl(} \\
 & &\!\! \mbox{} + \, d \, \Bigl( (-1)^{(r-1)(s-1)} \, i_Y^{} f \, - \,
                                  i_X^{} g \, - \,
                                  (-1)^{(r-1)s} \, i_Y^{} i_X^{} \theta \,
                                  \Bigr)~.
\end{eqnarray*}
In order for the bracket to be well defined, it is necessary and sufficient
that this expression vanishes whenever $X$ or $Y$ takes its values in the
kernel of $\omega$: this is guaranteed precisely when $f$ and $g$ are weak
Poisson forms, taking into account that $\theta$ is required to be a Poisson
form. Moreover, in view of eq.~(\ref{eq:SCHBRPB1}), eq.~(\ref{eq:POISB3})
follows immediately from eq.~(\ref{eq:POISB2}), proving that the Poisson
bracket $\{f,g\}$ of two weak Poisson forms is a Hamiltonian form. To check
that it is in fact a weak Poisson form and even a Poisson form when $f$ and
$g$ are Poisson forms, assume $\xi$ to be a multivector field taking values
in the kernel of $\,\omega$, say of degree $k$, and consider the expressions
obtained by contracting each of the four terms in eq.~(\ref{eq:POISB2}) with
$\xi$. The first obviously vanishes, whereas the fourth can be seen to vanish
due to eqs~(\ref{eq:KERNI}) and~(\ref{eq:KERN3}):
\begin{eqnarray*}
 i_\xi^{} \, d \, ( i_Y^{} i_X^{} \>\! \theta ) \!\!
 &=&\!\! (-1)^s \, i_\xi^{} \;\! i_Y^{} \, d \, ( i_X^{} \>\! \theta ) \, + \,
         i_\xi^{} \>\! L_Y^{} i_X^{} \>\! \theta \\
 &=&\!\! (-1)^{s(k-1)} \, i_Y^{} i_\xi^{} L_X^{} \>\! \theta \, + \,
         (-1)^{r+s(k-1)} \, i_Y^{} i_\xi^{} \>\! i_X^{} \, d \>\! \theta \\
 & &\!\! - \;  i_{[\>\!Y,\,\xi\,]}^{} i_X^{} \>\! \theta \,
         + \, (-1)^{(s-1)k} \, L_Y^{} i_\xi^{} \;\! i_X^{} \>\! \theta \\
 &=&\!\! - \; (-1)^{s(k-1)} \, i_Y^{} i_{[X,\,\xi\,]}^{} \>\! \theta \, + \,
         (-1)^{(r-1)k+s(k-1)} \, i_Y^{} L_X^{} i_\xi^{} \>\! \theta \\
 & &\!\! - \; (-1)^{r+s(k-1)} \, i_Y^{} i_\xi^{} \;\! i_X^{} \>\! \omega \\
 & &\!\! - \;  i_{[\>\!Y,\,\xi\,]}^{} i_X^{} \>\! \theta \, + \,
         (-1)^{(s-1)k} \, L_Y^{} i_\xi^{} \;\! i_X^{} \>\! \theta \\[1mm]
 &=&\!\! 0~.
\end{eqnarray*}
Similarly, the second and third can be handled by using eqs~(\ref{eq:KERNI})
and~(\ref{eq:KERN1}) which imply that
\begin{eqnarray*}
 i_\xi^{} \, d \, ( i_Y^{} f ) \!\!
 &=&\!\! (-1)^s \, i_\xi^{} \;\! i_Y^{} \, df \, + \,
         i_\xi^{} \>\! L_Y^{} f \\
 &=&\!\! (-1)^s \, i_\xi^{} \;\! i_Y^{} i_X^{} \>\! \omega \, - \,
         i_{[\>\!Y,\,\xi\,]}^{} f \, + \,
         (-1)^{(s-1)k} \, L_Y^{} i_\xi^{} f~,
\end{eqnarray*}
and
\begin{eqnarray*}
 i_\xi^{} \, d \, ( i_X^{} g ) \!\!
 &=&\!\! (-1)^r \, i_\xi^{} \;\! i_X^{} \, dg \, + \,
         i_\xi^{} \>\! L_X^{} g \\
 &=&\!\! (-1)^r \, i_\xi^{} \;\! i_X^{} i_Y^{} \>\! \omega \, - \,
         i_{[\>\!X,\,\xi\,]}^{} g \, + \,
         (-1)^{(r-1)k} \, L_X^{} i_\xi^{} g~.
\end{eqnarray*}
Indeed, $d$ applied to these expressions vanishes if $f$ and $g$ are weak
Poisson forms whereas the expressions themselves vanish if $f$ and $g$ are
Poisson forms.
\PCPqed

\noindent
Now we can formulate the main theorem of this paper:
\begin{thm}
 Let $P$ be an exact multisymplectic manifold. The Poisson bracket intro\-duced
 above is bilinear over $\mathbb{R}$, is graded antisymmetric, which means that
 for any two Poisson forms $f$ of degree $n-r$ and $g$ of degree $n-s$ on
 $P$, we have
 \begin{equation} \label{eq:POISBR12}
  \{g,f\}~=~- \, (-1)^{(r-1)(s-1)} \, \{f,g\}~,
 \end{equation}
 and satisfies the graded Jacobi identity, which means that for any three
 Poisson forms $f$ of degree $n-r$, $g$ of degree $n-s$ and $h$ of degree
 $n-t$ on $P$, we have
 \begin{equation} \label{eq:POISBR13}
  (-1)^{(r-1)(t-1)} \{f,\{g,h\}\} \; + \; \mathrm{cyclic~perm.}~=~0~,
 \end{equation}
 thus turning the space of Poisson forms on $P$ into a Lie superalgebra,
 The same statement holds if the term ``Poisson form'' is replaced by the
 term ``weak Poisson form''.
\end{thm}
\textbf{Remark}~~Bilinearity over $\mathbb{R}$ and the graded antisymmetry
(\ref{eq:POISBR12}) being obvious, the main statement of the theorem is of
course the validity of the graded Jacobi identity (\ref{eq:POISBR13}),
which depends crucially on the exact correction terms, that is, the last
three terms in the defining equation (\ref{eq:POISB2}). To prove this,
we need the following two lemmas:
\begin{lem}
 Let $P$ be a multisymplectic manifold. For any three locally Hamiltonian
 multivector fields $X$ of degree $r$, $Y$ of degree $s$ and $Z$ of degree
 $t$ on $P$, we have the cyclic identity
 \begin{equation} \label{eq:POISBR14}
  (-1)^{r(t-1)} \, i_X^{} \, d \, ( i_Y^{} i_Z^{} \>\! \omega ) \; + \;
  \mathrm{cyclic~perm.}~
  =~(-1)^{rt} \, d \, ( i_X^{} i_Y^{} i_Z^{} \>\! \omega )~,
 \end{equation}
\end{lem}
\textbf{Proof}~~This is obtained by calculating
\begin{eqnarray*}
 i_X^{} \, d \, ( i_Y^{} i_Z^{} \>\! \omega ) \!\!
 &=&\!\! (-1)^{(s-1)t} \, i_X^{} i_{[Y,Z]}^{} \>\! \omega~
  =~     (-1)^{(s-1)t+r(s+t-1)} \, i_{[Y,Z]}^{} i_X^{} \omega           \\[2mm]
 &=&\!\! (-1)^{r(s+t-1)} \,
         ( L_Y^{} i_Z^{} \, - \, (-1)^{(s-1)t} \, i_Z^{} L_Y^{} ) \,
         i_X \omega                                                     \\[2mm]
 &=&\!\! (-1)^{r(s+t-1)} \, d \, ( i_Y^{} i_Z^{} i_X^{} \omega ) \, + \;
         (-1)^{r(s+t-1)+s-1} \, i_Y^{} \, d \, ( i_Z^{} i_X^{} \omega ) \\[1mm]
 & &\!   \mbox{} - \, (-1)^{r(s+t-1)+(s-1)t} \,
         i_Z^{} \, d \, ( i_Y^{} i_X^{} \omega )~,
\end{eqnarray*}
and multiplying by $(-1)^{rt-r}$. \PCPqed
\begin{lem}
 Let $P$ be an exact multisymplectic manifold. For any three locally
 Hamiltonian multivector fields $X$ of degree $r$, $Y$ of degree $s$
 and $Z$ of degree $t$ on $P$, we have the cyclic identity
 \begin{equation} \label{eq:POISBR15}
  \begin{array}{l}
   (-1)^{r(t-1)} \, i_X^{} \, d \, ( i_Y^{} i_Z^{} \theta ) \, - \,
   (-1)^{r(t-1)+s} \, i_X^{} i_Y^{} \, d \, ( i_Z^{} \theta ) \; + \;
   \mathrm{cyclic~perm.} \\[2mm]
   \qquad =~(-1)^{rt+r+s+t} \, i_X^{} i_Y^{} i_Z^{} \>\! \omega \, + \,
            (-1)^{rt} \, d \left( i_X^{} i_Y^{} i_Z^{} \>\! \theta \right)\!~.
  \end{array}
 \end{equation}
\end{lem}
\textbf{Proof}~~This is obtained by calculating
\begin{eqnarray*}
\lefteqn{i_X^{} \, d \, ( i_Y^{} i_Z^{} \>\! \theta ) \, + \,
         (-1)^{s-1} \, i_X^{} i_Y^{} \, d \, ( i_Z^{} \>\! \theta )}
                                                          \hspace{-3mm} \\[1mm]
 &-&\!\! (-1)^{(s-1)t} \, i_X^{} i_Z^{} \, d \, ( i_Y^{} \theta ) \, + \,
         (-1)^{(s-1)(t-1)} \, i_X^{} i_Z^{} i_Y^{} \>\! \omega          \\[2mm]
 & & =~i_X^{} \, ( L_Y^{} i_Z^{} \, - \, (-1)^{(s-1)t} i_Z^{} L_Y^{} ) \,
       \theta                                                           \\[2mm]
 & & =~(-1)^{(s-1)t} \, i_X^{} i_{[Y,Z]}^{} \>\! \theta~
     = \;   (-1)^{(s-1)t+r(s+t-1)} \, i_{[Y,Z]}^{} i_X^{} \theta        \\[2mm]
 & & =~(-1)^{r(s+t-1)} \, ( L_Y^{} i_Z^{} \, - \, (-1)^{(s-1)t} \,
                            i_Z^{} L_Y^{} ) \, i_X^{} \theta            \\[2mm]
 & & =~(-1)^{r(s+t-1)} \, d \, ( i_Y^{} i_Z^{} i_X^{} \theta ) \, + \,
       (-1)^{r(s+t-1)+s-1} \, i_Y^{} \, d \, ( i_Z^{} i_X^{} \theta )   \\[1mm]
 & & \phantom{=~} \mbox{} - \,
       (-1)^{r(s+t-1)+(s-1)t} \, i_Z^{} \, d \, ( i_Y^{} i_X^{} \theta )
       \, - \,
       (-1)^{r(s+t-1)+(s-1)(t-1)} \, i_Z^{} i_Y^{} \, d \, ( i_X^{} \theta )~,
\end{eqnarray*}
and multiplying by $(-1)^{rt-r}$. \PCPqed

\vspace{4mm}

\noindent
\textbf{Proof of the theorem}~~Given any three weak Poisson forms $f$ of degree
$n-r$, $g$ of degree $n-s$ and $h$ of degree $n-t$ and fixing three Hamiltonian
multivector fields $X$ of degree $r$, $Y$ of degree $s$ and $Z$ of degree $t$
associated with $f$, with $g$ and with $h$, respectively, we compute the
double Poisson bracket
\vspace{2mm}
\begin{eqnarray*}
\lefteqn{(-1)^{(r-1)(t-1)} \, \{f,\{g,h\}\}}               \hspace{1cm} \\[3mm]
 &=&\!\! (-1)^{(r-1)(t-1) + r(s+t)} \,
         i_{[Z,Y]}^{} i_X^{} \omega                                     \\[1mm]
 & &\!   \mbox{} + \, (-1)^{(r-1)(t-1) + (r-1)(s+t)} \,
         d \, ( i_{[Z,Y]}^{} f )                                        \\[1mm]
 & &\!   \mbox{} - \, (-1)^{(r-1)(t-1)} \,
         d \, ( i_X^{} \{g,h\} )                                        \\[1mm]
 & &\!   \mbox{} - \, (-1)^{(r-1)(t-1) + (r-1)(s+t-1)} \,
         d \, ( i_{[Z,Y]}^{} i_X^{} \theta )                            \\[3mm]
 &=&\!   \mbox{} - \, (-1)^{(rs+r+t) + r(s+t-1) + (st+s+t)} \,
         i_X^{} i_{[\>\!Y,Z]}^{} \>\! \omega                            \\[2mm]
 & &\!   \mbox{} + \, (-1)^{(r-1)(s-1) + (t-1)s} \,
         d \, ( L_Z^{} i_Y^{} f ) \, - \,
         (-1)^{(r-1)(s-1)} \,
         d \, ( i_Y^{} L_Z^{} f )                                       \\[2mm]
 & &\!   \mbox{} - \, (-1)^{(r-1)(t-1) + s(t-1)} \,
         d \, ( i_X^{} i_Z^{} i_Y^{} \omega )                           \\[1mm]
 & &\!   \mbox{} - \, (-1)^{(r-1)(t-1) + (s-1)(t-1)} \,
         d \, ( i_X^{} \, d \, ( i_Z^{} g ) ) \, + \,
         (-1)^{(r-1)(t-1)} \,
         d \, ( i_X^{} \, d \, ( i_Y^{} h ) )                           \\[1mm]
 & &\!   \mbox{} + \, (-1)^{(r-1)(t-1) + (s-1)t} \,
         d \, ( i_X^{} \, d \, ( i_Z^{} i_Y^{} \theta ) )               \\[2mm]
 & &\!   \mbox{} - \, (-1)^{(r-1)s + (t-1)s} \,
         d \, ( L_Z^{} i_Y^{} i_X^{} \theta ) \, + \,
         (-1)^{(r-1)s} \,
         d \, ( i_Y^{} L_Z^{} i_X^{} \theta )
\end{eqnarray*}
\begin{eqnarray*}
 &=&\!   \mbox{} - \, (-1)^{rt+s+t} \,
         i_X^{} \, d \, ( i_Y^{} i_Z^{} \>\! \omega )                   \\[2mm]
 & &\!   \mbox{} + \, (-1)^{rs+st+r+t} \,
         d \, ( i_Z^{} \, d \, ( i_Y^{} f ) ) \, + \,
         (-1)^{rs+r+s} \,
         d \, ( i_Y^{} \, d \, ( i_Z^{} f ) )                           \\[1mm]
 & &\!   \mbox{} - \, \underline{(-1)^{rs+r+s+t} \,
         d \, ( i_Y^{} i_Z^{} i_X^{} \omega )}                          \\[2mm]
 & &\!   \mbox{} + \, \underline{(-1)^{rt+st+r+s+t} \,
         d \, ( i_X^{} i_Z^{} i_Y^{} \omega )}                          \\[1mm]
 & &\!   \mbox{} - \, (-1)^{rt+st+r+s} \,
         d \, ( i_X^{} \, d \, ( i_Z^{} g ) ) \, - \,
         (-1)^{rt+r+t} \,
         d \, ( i_X^{} \, d \, ( i_Y^{} h ) )                           \\[1mm]
 & &\!   \mbox{} - \, (-1)^{rt+r} \,
         d \, ( i_X^{} \, d \, ( i_Y^{} i_Z^{} \theta ) )
         \quad \leftarrow                                               \\[1mm]
 & &\!   \mbox{} + \, (-1)^{st+t} \,
         d \, ( i_Z^{} \, d \, ( i_X^{} i_Y^{} \theta ) )
         \quad \leftarrow                                               \\[1mm]
 & &\!   \mbox{} + \, (-1)^{rs+s} \,
         d \, ( i_Y^{} \, d \, ( i_Z^{} i_X^{} \theta ) ) \, - \,
         (-1)^{rs+s+t} \,
         d \, ( i_Y^{} i_Z^{} \, d \, ( i_X^{} \theta ) )~.
\vspace{5mm}
\end{eqnarray*}
In the last expression, the underlined terms cancel each other. Moreover,
under the cyclic sum, the terms marked by an arrow cancel each other and
the terms containing derivatives of contractions of $f$, $g$, $h$ cancel
pairwise, i.e., the expression
\begin{eqnarray*}
 &+&\!\!\! (-1)^{rs+st+r+t} \,
           d \, ( i_Z^{} \, d \, ( i_Y^{} f ) ) \, + \,
           (-1)^{rs+r+s} \,
           d \, ( i_Y^{} \, d \, ( i_Z^{} f ) )                              \\
 &-&\!\!\! (-1)^{rt+st+r+s} \,
           d \, ( i_X^{} \, d \, ( i_Z^{} g ) ) \, - \,
           (-1)^{rt+r+t} \,
           d \, ( i_X^{} \, d \, ( i_Y^{} h ) )                         \\[1mm]
 &+&\!\!\! (-1)^{st+tr+s+r} \,
           d \, ( i_X^{} \, d \, ( i_Z^{} g ) ) \, + \,
           (-1)^{st+s+t} \,
           d \, ( i_Z^{} \, d \, ( i_X^{} g ) )                              \\
 &-&\!\!\! (-1)^{sr+tr+s+t} \,
           d \, ( i_Y^{} \, d \, ( i_X^{} h ) ) \, - \,
           (-1)^{sr+s+r} \,
           d \, ( i_Y^{} \, d \, ( i_Z^{} f ) )                         \\[1mm]
 &+&\!\!\! (-1)^{tr+rs+t+s} \,
           d \, ( i_Y^{} \, d \, ( i_X^{} h ) ) \, + \,
           (-1)^{tr+t+r} \,
           d \, ( i_X^{} \, d \, ( i_Y^{} h ) )                              \\
 &-&\!\!\! (-1)^{ts+rs+t+r} \,
           d \, ( i_Z^{} \, d \, ( i_Y^{} f ) ) \, - \,
           (-1)^{ts+t+s} \,
           d \, ( i_Z^{} \, d \, ( i_X^{} g ) )
\end{eqnarray*}
vanishes. Finally, using the cyclic identities (\ref{eq:POISBR14}) and
(\ref{eq:POISBR15}), we see that the remaining terms sum up as follows:
\begin{eqnarray*}
\lefteqn{(-1)^{(r-1)(t-1)} \, \{f,\{g,h\}\} \; + \; \mathrm{cyclic~perm.}}
                                                           \hspace{5mm} \\[3mm]
 &=&\!\! - \, (-1)^{r+s+t} \Bigl(
 (-1)^{r(t-1)} \, i_X \, d \, ( i_Y^{} i_Z^{} \omega ) \; + \;
                \mathrm{cyclic~perm.} \Bigr)                            \\[1mm]
 & &\!\! + \, d \, \Bigl(
           (-1)^{r(t-1)} \, i_X^{} \, d \, ( i_Y^{} i_Z^{} \theta ) \, - \,
           (-1)^{r(t-1)+s} \, i_X^{} i_Y^{} \, d \, ( i_Z^{} \theta ) \; + \;
           \mathrm{cyclic~perm.} \Bigr)                                 \\[2mm]
 &=&\!\! - \, (-1)^{r+s+t} \, (-1)^{rt} \,
         d \, ( i_X^{} i_Y^{} i_Z^{} \omega )                           \\[1mm]
 & &\!\! + \,  d \, \Bigl(
         (-1)^{r+s+t} \, (-1)^{rt} \, i_X^{} i_Y^{} i_Z^{} \omega \, + \,
         (-1)^{rt} \, d \, ( i_X^{} i_Y^{} i_Z^{} \theta ) \Bigr)       \\[3mm]
 &=&\!\! 0~.
\end{eqnarray*}
This completes the proof of the main theorem. \PCPqed

\noindent
\textbf{Remark}~~From the definition given in eq.~(\ref{eq:POISB2}), it is
obvious that the Poisson bracket between an arbitrary (weak) Poisson form
$f$ and a closed (weak) Poisson form $g$ is exact, since in this case the
Hamiltonian multivector field $Y$ associated with $g$ may be chosen to
vanish identically, so that one gets $\, \{f,g\} = - \, d \, ( i_X^{} g )$.
Therefore, the space of closed (weak) Poisson forms is an ideal in the Lie
superalgebra of all (weak) Poisson forms.

Concluding, it must not go unnoticed that the Poisson bracket between
Poisson forms introduced in this paper should be looked upon with a
certain amount of caution, for a variety of reasons. One of these is
that the space of Poisson forms is a Lie superalgebra but apparently
not a Poisson superalgebra, since the Poisson bracket does not act as
a superderivation in its second argument with respect to the exterior
product of forms, nor does there seem to exist any other naturally
defined associative supercommutative product between Poisson forms
with that property: this is in contrast to the situation for multi%
vector fields which do form a Poisson superalgebra with respect to
the exterior product and the Schouten bracket. There is also a degree
problem, since for example, the Poisson bracket between functions would
be a form of negative degree, which is always zero: this is, at least
at first sight, rather odd. Finally, the question about the relation
to the covariant Poisson bracket of Peierls and de Witt mentioned at
the end of the introduction remains open.


\section{The universal multimomentum map}

On exact multisymplectic manifolds, Definition \ref{dfn:HAMMVFF} can be
complemented as follows.
\begin{dfn} \label{dfn:EHAMMVF}
 A multivector field $X$ on an exact multisymplectic manifold $P$
 is called \textbf{exact Hamiltonian\/} if
 \begin{equation} \label{eq:EHMVF}
  L_X^{} \theta~=~0~.
 \end{equation}
\end{dfn}
The terminology is consistent with that introduced before because exact
Hamiltonian multivector fields are Hamiltonian: this is an immediate
consequence of Proposition~\ref{prp:UNMMM1} below. Thus Proposition
\ref{prp:HAMMVFF} can be complemented as follows.
\begin{prp} \label{prp:EHAMMVF}
 The Schouten bracket of any two exact Hamiltonian multivector fields
 $X$ and $Y$ on an exact multisymplectic manifold $P$ is an exact
 Hamiltonian multivector field $\, [X,Y]$ on $P$. This means that
 the space $\mathfrak{X}_{EH}^\wedge(P)$ of exact Hamiltonian
 multivector fields on $P$ is a subalgebra of the Lie superalgebra
 $\mathfrak{X}^\wedge(P)$ of all multivector fields on~$P$ which,
 according to eq.~(\ref{eq:KERNI}), contains the space
 $\mathfrak{X}_0^\wedge(P)$ of multivector fields taking
 values in the kernel of $\,\omega$ as an ideal.
\end{prp}
\textbf{Proof}~~The proposition follows directly from eq.~(\ref{eq:LXLY-LYLX}).
\PCPqed

\noindent
Exact Hamiltonian multivector fields generate Poisson forms, by contraction
with the multicanonical form.
\begin{prp} \label{prp:UNMMM1}
 Let $P$ be an exact multisymplectic manifold. For every exact Hamiltonian
 $r$-multivector field $X$ on $P$, the formula
 \begin{equation} \label{eq:UNMMM1}
  J(X)~=~(-1)^{r-1} \, i_X^{} \theta
 \end{equation}
 defines a Poisson $(n-r)$-form $J(X)$ on $P$ whose associated Hamiltonian
 multi\-vector field is $X$ itself. In particular, $X$ is Hamiltonian.
\end{prp}
\textbf{Proof}~~Using eq.~(\ref{eq:LDFMVF}), we see that the condition
(\ref{eq:EHMVF}) implies
\begin{equation} \label{eq:UNMMM2}
 d \left( J(X) \right)\!~=~(-1)^{r-1} \, d \left( i_X^{} \theta \right)\!~
 =~(-1)^{r-1} \, L_X^{} \theta \, - \, i_X^{} \, d \>\! \theta~
 =~i_X^{} \omega~,
\end{equation}
so $J(X)$ is a Hamiltonian form whose associated Hamiltonian multivector
field is $X$ itself. Moreover, the kernel of $J(X)$ on multivectors
contains that of $\theta$ which in turn contains that of $\omega$,
so $J(X)$ is a Poisson form. \PCPqed
\begin{prp} \label{prp:UNMMM2}
 Let $P$ be an exact multisymplectic manifold. The linear map~$J$
 from the space $\mathfrak{X}_{EH}^\wedge(P)$ of exact Hamiltonian
 multivector fields on~$P$ to the space of Poisson forms on~$P$
 defined by eq.~(\ref{eq:UNMMM1}) is an antihomomorphsim of Lie
 superalgebras, i.e., we have
 \begin{equation} \label{eq:UNMMM3}
  \{J(X),J(Y)\}~=~J([\>\!Y,X])~.
 \end{equation}
\end{prp}
\textbf{Proof}~~For any two exact Hamiltonian multivector fields $X$
of degree~$r$ and $Y$ of degree~$s$, we have, according to the defining
equations~(\ref{eq:POISB2}) and~(\ref{eq:UNMMM1}),
\begin{eqnarray*}
 \{J(X),J(Y)\} \!\!
 &=&\!\! (-1)^{r(s-1)} \, i_Y^{} i_X^{} \omega \, + \,
         (-1)^{(r-1)(s-1)+r-1} \, d \left( i_Y^{} i_X^{} \theta \right)      \\
 & &\!\! \mbox{}
         - \, (-1)^{s-1} \, d \left( i_X^{} i_Y^{} \theta \right) \,
         - \, (-1)^{(r-1)s} \, d \left( i_Y^{} i_X^{} \theta \right)    \\[1mm]
 &=&\!\! (-1)^{r(s-1)} \, i_Y^{} i_X^{} \omega \, + \,
         (-1)^{(r-1)s} \, d \left( i_Y^{} i_X^{} \theta \right)~,
\end{eqnarray*}
whereas combining eqs~(\ref{eq:LXiY-iYLX}), (\ref{eq:LDFMVF}) and
(\ref{eq:UNMMM2}) gives
\begin{eqnarray*}
 J([\>\!Y,X]) \!\!
 &=&\!\! (-1)^{r+s} \, i_{[\>\!Y,X]}^{} \>\! \theta                     \\[1mm]
 &=&\!\! (-1)^{r+s+r(s-1)} \, L_Y^{} i_X^{} \theta \qquad
 \mbox{since $\, L_Y^{} \theta = 0$}                                    \\[1mm]
 &=&\!\! (-1)^{r(s-1)} \, d \left( i_Y^{} i_X^{} \theta \right) \, - \,
         (-1)^{r(s-1)+s} i_Y^{} \, d \left( i_X^{} \theta \right)       \\[1mm]
 &=&\!\! (-1)^{r(s-1)} \, d \left( i_Y^{} i_X^{} \theta \right) \, + \,
         (-1)^{r(s-1)} i_Y^{} i_X^{} \omega~.
\end{eqnarray*}
Obviously, these two expressions coincide. \PCPqed

\noindent
\textbf{Remark}~~This proposition, even when restricted to vector fields and
$(n\!-\!1)$-forms, constitutes a remarkable improvement over the corresponding
Proposition 4.5 of Ref.~\cite{GIM} where, due to an inadequate definition
of the Poisson bracket (omitting the exact correction terms, that is, the
last three terms in eq.~(\ref{eq:POISB2})), eq.~(\ref{eq:UNMMM3}) must be
modified by an exact correction term.
\begin{dfn} \label{dfn:UNMMM}
 Let $P$ be an exact multisymplectic manifold. The linear map~$J$ from the
 space $\mathfrak{X}_{EH}^\wedge(P)$ of exact Hamiltonian multivector fields
 on~$P$ to the space of Poisson forms on~$P$ defined by eq.~(\ref{eq:UNMMM1})
 will be called the \textbf{universal multimomentum map} and its restriction
 to the space $\mathfrak{X}_{EH}(P)$ of exact Hamiltonian vector fields on~$P$
 the \textbf{universal momentum map}.
\end{dfn}
\textbf{Remark}~~The term ``universal momentum map'' can be justified
in the context of Noether's theorem, dealing with the derivation of
conservation laws from symmetries. In classical field theory, conserved
quantities are described by Noether currents which depend on the fields
of the theory and are $(n\!-\!1)$-forms on $n$-dimensional space-time,
so that they can be integrated over compact regions in spacelike hyper%
surfaces in order to provide Noether charges associated with each such
region: Noether's theorem then asserts that when the fields satisfy the
equations of motion of the theory, these Noether currents are closed forms.
In the multiphase space approach, the Noether currents on space-time are
obtained from corresponding Noether current forms defined on (extended)
multiphase space via pull-back of differential forms, their entire field
dependence being induced by this pull-back. Moreover, there is an explicit
procedure to construct these Noether current forms on (extended) multiphase
space: it is the field theoretical analogue of the momentum map of
Hamiltonian mechanics on cotangent bundles and, in Ref.~\cite{GIM},
is called the ``special covariant momentum map''. Briefly, given a
Lie group~$G$, with Lie algebra~$\mathfrak{g}$, the statement that~$G$
is a symmetry group of a specific theory supposes that we are given an
action of~$G$ on the configuration bundle~$E$ over~$M$ by bundle
automorphisms, which of course induces actions of~$G$ on~$JE$
and on~$\vec{J} E$, as well as on all of their duals, including
$\vec{J}\upoast E$ and $J\upostar E$, by bundle automorphisms. \linebreak
(In order to speak of a symmetry, we must also assume the Lagrangian or
Hamiltonian density to be invariant, or rather equivariant, under the
action of~$G$, but this aspect is not relevant for the present discussion.)
As usual, each of these actions induces an antihomomorphism from~%
$\mathfrak{g}$ to the Lie algebra of vector fields on the corresponding
manifold, taking each generator~$X$ in~$\mathfrak{g}$ to the corresponding
fundamental vector field $X_M^{}$, $X_E^{}$, $X_{JE}^{}$, $X_{\vec{J} E}$
$\ldots$ $X_{\vec{J}\smupoast E}$, $X_{J\smupostar E}^{}$, all of which
(except $X_M^{}$) are projectable: for example, $X_E^{}$ projects to
$X_M^{}$ under the tangent map $\, T\pi : TE \rightarrow TM \,$ to the
projection $\, \pi : E \rightarrow M$. Moreover, the vector fields
$X_{JE}^{}$, $X_{\vec{J} E}$ $\ldots$ $X_{\vec{J}\smupoast E}$,
$X_{J\smupostar E}^{}$ can all be obtained from the vector field~%
$X_E^{}$ by a canonical lifting process. In particular, the projectable
vector fields~$X_{J\smupostar E}$ on~$J\upostar E$ obtained from projectable
vector fields~$X_E$ on~$E$ by lifting are exact Hamiltonian, and conversely,
it turns out that all exact Hamiltonian vector fields on $J\upostar E$ are
obtained in this way. (The last statement, analogous to a corresponding
statement for cotangent bundles, is not proved in Ref.~\cite{GIM}; it
will be derived in Ref.\ \cite{FPR2}.)
Now the ``special covariant momentum map'' of Ref.~\cite{GIM}
associated with the symmetry under~$G$ is simply given by composing the
antihomomorphism that takes generators~$X$ in~$\mathfrak{g}$ to exact
Hamiltonian fundamental vector fields~$X_{J\smupostar E}^{}$ on~$J\upostar E$
with the universal momentum map introduced above. Therefore, the universal
momentum map comprises that part of the construction of the momentum map
in field theory which does not depend on the \emph{a priori\/} choice of
a symmetry group or its action on the dynamical variables of the theory,
and the universal multimomentum map extends that from vector fields to
multivector fields.


\section{Poisson forms on multiphase space}

Our aim in this final section is to give a series of examples for Poisson
forms on the extended multiphase space $J\upostar E$ of field theory.
A full, systematic treatment of the subject will be given in a forth%
coming separate paper~\cite{FPR2}.

As a preliminary step, we observe that there is a natural, globally defined
notion of vertical vectors and of horizontal covectors on $J\upostar E$.
In fact, there are two such~notions, one referring to the ``source''
projection onto space-time $M$ and the other to the ``target'' projection
onto the total space $E$ of the configuration bundle. In either case, the
vertical vectors are those that vanish under the tangent to the projection,
while the horizontal covectors are those that vanish on all vertical vectors.
In adapted local coordinates,
\begin{eqnarray}
 &{\displaystyle
   \frac{\partial}{\partial q^i}~~,~~
   \frac{\partial}{\partial p\>\!_i^\mu}
   \quad \mbox{and} \quad
   \frac{\partial}{\partial p}
   \qquad \mbox{are vertical with respect to the source projection}~,\quad}& \\
 &{\displaystyle
   \frac{\partial}{\partial p\>\!_i^\mu}
   \quad \mbox{and} \quad
   \frac{\partial}{\partial p}
   \qquad \mbox{are vertical with respect to the target projection}~,\quad}&
\end{eqnarray}
while
\vspace{-1mm}
\begin{eqnarray}
 &dx^\mu \qquad
  \mbox{are horizontal with respect to the source projection}~,\quad &  \\[3mm]
 &dx^\mu \quad \mbox{and} \quad dq^i \qquad
  \mbox{are horizontal with respect to the target projection}~.\quad &
\vspace{1mm}
\end{eqnarray}
This can be extended to multivectors and exterior forms, as follows. Given
positive integers $r$ and $s$ with $\, s\leqslant r$, an exterior $r$-form
is said to be $s$-horizontal if it vanishes whenever one inserts at least
$r-\!s+1$ vertical vectors (this includes the standard notion of horizontal
forms by taking $\, s=r$), and an $r$-multivector is said to be $s$-vertical
if it is annihilated by all $(r-\!s+1)$-horizontal exterior forms. Using the
standard expansion of multivectors and of exterior forms in adapted local
coordinates, it is not difficult to see that an $r$-form is $s$-horizontal
if and only if it is a linear combination of terms each of which is an
exterior product containing at least $s$ horizontal covectors and that an
$r$-multivector is $s$-vertical if and only if it is a linear combination of
terms each of which is an exterior product containing at least $s$ vertical
vectors. Thus for example, eqs.~(\ref{eq:MCANF02}), (\ref{eq:MSYMF02})
and~(\ref{eq:SVFEMPS}) show that $\theta$ and $\omega$ are both
$(n\!-\!1)$-horizontal with respect to the source projection and
even $n$-horizontal with respect to the target projection, while
$\Sigma$ is vertical with respect to both projections.

In what follows, the terms ``vertical'' and ``horizontal'' will always
refer to the source projection, except when explicitly stated otherwise.

For later use, we first write down the expansion of a general multivector
field $X$ of degree $r$ in terms of adapted local coordinates, as follows:
\begin{eqnarray} \label{eq:HAMMVF1}
 X \!\!
 &=&\!\! \frac{1}{r!} \; X_{\phantom{i}}^{\mu_1 \ldots\, \mu_r} \;
         \frac{\partial}{\partial x_{\phantom{i}}^{\mu_1}}
         \;\smwedge \ldots \smwedge\;
         \frac{\partial}{\partial x_{\phantom{i}}^{\mu_r}}    \nonumber \\[1mm]
 & &\!   \mbox{} + \; \frac{1}{(r\!-\!1)!} \;
         X_{\phantom{i}}^{i,\mu_2 \ldots\, \mu_r} \;
         \frac{\partial}{\partial q^i} \;\smwedge\;
         \frac{\partial}{\partial x_{\phantom{i}}^{\mu_2}}
         \;\smwedge \ldots \smwedge\;
         \frac{\partial}{\partial x_{\phantom{i}}^{\mu_r}}    \nonumber \\
 & &\!   \mbox{} +~~~~\frac{1}{r!}~~~~
         X_i^{\mu_1 \ldots\, \mu_r} \;
         \frac{\partial}{\partial p\>\!_i^{\mu_1}} \;\smwedge\;
         \frac{\partial}{\partial x_{\phantom{i}}^{\mu_2}}
         \;\smwedge \ldots \smwedge\;
         \frac{\partial}{\partial x_{\phantom{i}}^{\mu_r}}              \\
 & &\!   \mbox{} + \; \frac{1}{(r\!-\!1)!} \; X_0^{\mu_2 \ldots\, \mu_r} \;
         \frac{\partial}{\partial p} \;\smwedge\;
         \frac{\partial}{\partial x_{\phantom{i}}^{\mu_2}}
         \;\smwedge \ldots \smwedge\;
         \frac{\partial}{\partial x_{\phantom{i}}^{\mu_r}}    \nonumber \\[3mm]
 & &\!   \mbox{} + \; \xi~.                                   \nonumber
\end{eqnarray}
Here, all coefficients are assumed to be totally antisymmetric in their
space-time indices, whereas $\xi$ is assumed to take values in the kernel
of $\omega$. (This can always be achieved without loss of generality,
because if we begin by supposing instead that $\xi$ should contain all
other terms of the standard expansion, that is, all $2$-vertical terms,
then $\xi$ would contain just one group of terms that are not obviously
annihilated under contraction with $\omega$, namely the terms of the form
\[
 \frac{\partial}{\partial q^i} \;\smwedge\;
 \frac{\partial}{\partial p\>\!_k^\kappa} \;\smwedge\;
 \frac{\partial}{\partial x^{\mu_3}} \;\smwedge \ldots \smwedge\;
 \frac{\partial}{\partial x^{\mu_r}}~.
\]
However, this part of $\xi$ can be decomposed into the sum of a term which is
annihilated under contraction with $\omega$ and a linear combination of the
$1$-vertical terms
\[
 \frac{\partial}{\partial p} \;\smwedge\;
 \frac{\partial}{\partial x_{\phantom{i}}^{\mu_2}} \;\smwedge\;
 \frac{\partial}{\partial x_{\phantom{i}}^{\mu_3}}
 \;\smwedge \ldots \smwedge\;
 \frac{\partial}{\partial x_{\phantom{i}}^{\mu_r}}~,
\]
so that by a redefinition of the coefficents $X_0^{\mu_2 \ldots\, \mu_r}$ and
of $\xi$, we arrive at the expression for~$X$ given in eq.~(\ref{eq:HAMMVF1}),
with $\xi$ now taking values in the kernel of $\omega$. For a more detailed
discussion, see Ref.~\cite{FPR2}.) Explicitly, the contraction of $\omega$
with $X$ then reads
\begin{equation} \label{eq:HAMMVF2}
 \begin{array}{rcl}
  i_X^{} \omega \!\!
  &=&\!\! {\displaystyle \frac{1}{r!} \; X^{\mu_1 \ldots\, \mu_r} \;
           dq^i \,\smwedge\; dp\>\!_i^\mu \,\smwedge\;
           d^{\,n} x_{\mu \mu_1 \ldots\, \mu_r} \; - \;
           \frac{(-1)^r}{r!} \; X^{\mu_1 \ldots\, \mu_r} \;
           dp \;\smwedge\; d^{\,n} x_{\mu_1 \ldots\, \mu_r}}            \\[4mm]
  & &\!   {\displaystyle \mbox{} + \;
           \frac{(-1)^{r-1}}{(r\!-\!1)!} \; X^{i,\mu_2 \ldots\, \mu_r} \;
           dp\>\!_i^\mu \,\smwedge\;
           d^{\,n} x_{\mu \mu_2 \ldots\, \mu_r}}                        \\[4mm]
  & &\!   {\displaystyle \mbox{} +~~
           \frac{(-1)^r}{r!}~~X_i^{\mu_1 \ldots\, \mu_r} \;
           dq^i \,\smwedge\; d^{\,n} x_{\mu_1 \ldots\, \mu_r}}          \\[4mm]
  & &\!   {\displaystyle \mbox{} - \;
           \frac{1}{(r\!-\!1)!}~X_0^{\mu_2 \ldots\, \mu_r} \;
           d^{\,n} x_{\mu_2 \ldots\, \mu_r}}~,
 \end{array}
\end{equation}
while that of $\theta$ with $X$ reads
\vspace{1mm}
\begin{equation} \label{eq:HAMMVF3}
 \begin{array}{rcl} 
  i_X^{} \theta \!\!
  &=&\!\! {\displaystyle
           \frac{(-1)^r}{r!} \; X^{\mu_1 \ldots\, \mu_r} \; p\>\!_i^\mu \;
           dq^i \,\smwedge\; d^{\,n} x_{\mu \mu_1 \ldots\, \mu_r} \; + \;
           \frac{1}{r!} \; X^{\mu_1 \ldots\, \mu_r} \; p \>\,
           d^{\,n} x_{\mu_1 \ldots\, \mu_r}}                            \\[4mm]
  & &\!   {\displaystyle
           \mbox{} + \; \frac{1}{(r\!-\!1)!} \; X^{i,\mu_2 \ldots\, \mu_r} \;
           p\>\!_i^\mu \, d^{\,n} x_{\mu \mu_2 \ldots\, \mu_r}}~,
 \end{array}
\end{equation}
where, in each of the last two equations, the first term is to be omitted
if $\, r=n$, whereas only the last term in the first equation remains and
$i_X^{} \theta$ vanishes identically if $\, r=n+1$.

With these preliminaries out of the way, we can easily deal with the simplest
case, which is that of functions.
\begin{prp}
 A function $f$ on $J\upostar E$ is always a Poisson $0$-form. Moreover,
 in adapted local coordinates, the corresponding Hamiltonian $n$-multivector
 field $X$ is, \mbox{modulo} terms taking values in the kernel of $\,\omega$,
 given by
 \vspace{2mm}
 \begin{equation} \label{eq:HAMNVF1}
  \begin{array}{rcl}
   X &=& {\displaystyle - \;
          \frac{1}{(n\!-\!1)!} \; \epsilon^{\mu_2 \ldots\, \mu_n \mu} \;
          \left( \frac{\partial f}{\partial x_{\phantom{i}}^\mu} \,
                 \frac{\partial}{\partial p} \; - \;
                 \frac{1}{n} \, \frac{\partial f}{\partial p} \,
                 \frac{\partial}{\partial x_{\phantom{i}}^\mu} \right)
          \,\smwedge\;
          \frac{\partial}{\partial x_{\phantom{i}}^{\mu_2}}
          \;\smwedge \ldots \smwedge\;
          \frac{\partial}{\partial x_{\phantom{i}}^{\mu_n}}} \\[5mm]
     & & {\displaystyle + \;
          \frac{1}{(n\!-\!1)!} \; \epsilon^{\mu_2 \ldots\, \mu_n \mu} \;
          \left( \frac{\partial f}{\partial p\>\!_i^\mu} \,
                 \frac{\partial}{\partial q^i} \; - \;
                 \frac{1}{n} \, \frac{\partial f}{\partial q^i} \,
                 \frac{\partial}{\partial p\>\!_i^\mu} \right)
          \,\smwedge\;
          \frac{\partial}{\partial x_{\phantom{i}}^{\mu_2}}
          \;\smwedge \ldots \smwedge\;
          \frac{\partial}{\partial x_{\phantom{i}}^{\mu_n}}}~.
  \end{array}
 \end{equation}
\end{prp}
\textbf{Proof}~~First of all, observe that for functions $f$, the kernel
condition (\ref{eq:KERN1}) is void. Next, we simplify the expression
(\ref{eq:HAMMVF2}), with $\, r=n$, by noting that due to our conventions
(\ref{eq:VOLDF1}), (\ref{eq:CMPDF1}) and~(\ref{eq:CMPDF2}), we have
\begin{equation} \label{eq:FULCTR1}
 d^{\,n} x_{\mu_1 \ldots\, \mu_n}~=~\epsilon_{\mu_1 \ldots\, \mu_n}~~~,~~~
 d^{\,n} x_{\mu_2 \ldots\, \mu_n}~
 =~\epsilon_{\mu_2 \ldots\, \mu_n \mu} \, dx^\mu~.
\end{equation}
Thus
\begin{equation} \label{eq:HAMNVF2}
 \begin{array}{rcl}
  i_X^{} \omega \!\!
  &=&\!\! {\displaystyle - \;
           \frac{(-1)^n}{n!} \; \epsilon_{\mu_1 \ldots\, \mu_n} \;
           X^{\mu_1 \ldots\, \mu_n} \; dp \; + \;
           \frac{1}{(n\!-\!1)!} \; \epsilon_{\mu_2 \ldots\, \mu_n \mu} \;
           X^{i,\mu_2 \ldots\, \mu_n} \; dp\>\!_i^\mu}                  \\[4mm]
  & &\!\! {\displaystyle - \;
           \frac{1}{(n\!-\!1)!} \; \epsilon_{\mu_2 \ldots\, \mu_n \mu} \;
           X_i^{\mu , \mu_2 \ldots\, \mu_n} \; dq^i \; - \;
           \frac{1}{(n\!-\!1)!} \; \epsilon_{\mu_2 \ldots\, \mu_n \mu} \;
           X_0^{\mu_2 \ldots\, \mu_n} \; dx^\mu}~.
 \end{array}
\end{equation}
Equating this expression with the exterior derivative of $f$, we obtain the
following system of equations
\begin{eqnarray}
 &X^{\mu_1 \ldots\, \mu_n}~
  =~(-1)^{n-1} \, \epsilon^{\mu_1 \ldots\, \mu_n} \,
    {\displaystyle \frac{\partial f}{\partial p}}~,&
 \label{eq:POIS0F1} \\[1mm]
 &X^{i,\mu_2 \ldots\, \mu_n}~
  =~\epsilon^{\mu_2 \ldots\, \mu_n \mu} \,
    {\displaystyle
     \frac{\partial f}{\partial p\>\!_i^\mu}}~,&
 \label{eq:POIS0F2} \\[1mm]
 &X_i^{\mu , \mu_2 \ldots\, \mu_n}~
  =~- \; \epsilon^{\mu_2 \ldots\, \mu_n \mu} \,
         {\displaystyle
         \frac{1}{n} \frac{\partial f}{\partial q^i}}~,&
 \label{eq:POIS0F3} \\[1mm]
 &X_0^{\mu_2 \ldots\, \mu_n}~
  =~- \; \epsilon^{\mu_2 \ldots\, \mu_n \mu} \,
         {\displaystyle
          \frac{\partial f}{\partial x^\mu}}~.& \rule[-4mm]{0mm}{6mm}
 \label{eq:POIS0F4}
\end{eqnarray}
Inserting this back into eq.~(\ref{eq:HAMMVF1}), with $\, r=n$, and
rearranging the terms, we arrive at eq.~(\ref{eq:HAMNVF1}).
\PCPqed

\noindent
\textbf{Remark}~~It has been shown in Ref.~\cite{PR} that for functions $h$
on $J\upostar E$ of the special form
\begin{equation} \label{eq:HAMFUN}
  h(x_{\phantom{i}}^\mu,q^i,p^\mu_i,p)~
  =~- \, H(x_{\phantom{i}}^\mu,q^i,p^\mu_i) \, - \, p~,
\end{equation}
the associated Hamiltonian multivector field $X$ can be chosen so that
it defines an $n$-dimensional distribution in $J\upostar E$ because it
is locally decomposable, that is, locally there exist vector fields
$\, X_1, \ldots , X_n \,$ such that $\, X = X_1 \;\smwedge \ldots
\smwedge\, X_n \,$ satisfies the equation $\, i_X^{} \omega = dh$.
Indeed, setting
\begin{equation} \label{eq:HAMNVF5}
 X_\mu~=~- \, \frac{\partial}{\partial x_{\phantom{i}}^\mu} \,
         + \, \frac{\partial h}{\partial p\>\!_i^\mu} \,
              \frac{\partial}{\partial q^i} \,
         - \, \frac{1}{n} \, \frac{\partial h}{\partial q^i} \,
              \frac{\partial}{\partial p\>\!_i^\mu} \,
         - \left( \frac{\partial h}{\partial x_{\phantom{i}}^\mu} \, - \,
                  \frac{1}{n} \, \frac{\partial h}{\partial q^i} \,
                  \frac{\partial h}{\partial p\>\!_i^\mu} \right)
           \frac{\partial}{\partial p}~,
\end{equation}
we can convince ourselves that this choice of $X$ and the choice of $X$ made
in eq.~(\ref{eq:HAMNVF1}) differ by a term taking values in the kernel of
$\omega$. Under additional assumptions, this distribution will be integrable
and its integral manifolds will be the images of \linebreak sections of
$J\upostar E$ over $M$ satisfying the covariant Hamiltonian equations of
motion, or \mbox{De Donder - Weyl} equations.

Another method for constructing Poisson forms on the extended multiphase
space $J\upostar E$ is from Hamiltonian forms on the ordinary multiphase 
space $\vec{J}\upoast E$, as introduced by Kanatchikov \cite{Ka1,Ka2},
pulling these back to $J\upostar E$ via the appropriate projection.

To describe the salient features of Kanatchikov's construction, one must
first of all introduce a structure on $\vec{J}\upoast E$ similar to the
multisymplectic form $\omega$ that exists naturally on $J\upostar E$.
This requires the choice of a connection in $E$ and of a linear
connection in $TM$ which, for the sake of convenience, will be
assumed to be torsion free. Together, they induce connections in all
the other bundles that are important in the multiphase space approach 
to field theory, including the multiphase spaces $\vec{J}\upoast E$
and $J\upostar E$; for the convenience of the reader, the relevant
formulas in adapted local coordinates are collected in Appendix B.
In the case of $\vec{J}\upoast E$, this induced connection can be
used to define a ''vertical multisymplectic form'' $\omega^V$ which
is however not closed; instead, it is annihilated under the action
of a ''vertical exterior derivative'' $d^{\,V}$ for differential
forms. In adapted local coordinates, these objects can be written
in the form
\begin{equation}
 \omega^V~=~e^i \,\smwedge\; e_i^\mu \,\smwedge\; d^{\,n} x_\mu^{} \,
            + \, \ldots
\end{equation}
and
\begin{equation}
 d^{\,V}~=~e^i \,\smwedge\; \frac{\partial}{\partial q^i} \, + \,
           e_i^\mu \,\smwedge\; \frac{\partial}{\partial p\>\!_i^\mu}
\end{equation}
respectively, where $\, e^i = dq^i + \Gamma_\nu^i \, dx^\nu \,$ and
$\, e_i^\mu = dp\>\!_i^\mu - \left( \partial_i^{} \Gamma_\kappa^j \,
p\>\!_j^\mu - \Gamma_{\kappa\lambda}^\mu p\>\!_i^\lambda +
\Gamma_{\kappa\rho}^\rho p\>\!_i^\mu \right) dx^\kappa$
are vertical $1$-forms (with respect to the aforementioned induced connection):
the dots in the definition of $\omega^V$ indicate $n$-horizontal terms that
are not important here, while the partial derivatives in the definition of
$d^{\,V}$ are meant to act on the coefficient functions. As shown by one of
the present authors \cite{Pau}, $d^{\,V}$ is still a cohomology operator,
i.e., it has square zero. Then the Hamiltonian forms as defined by
Kanatchikov can be shown to be precisely the horizontal forms
$\tilde{f}$ on $\vec{J}\upoast E$ satisfying the equation
\begin{equation} \label{eq:KAN1}
 i_{\tilde{X}}^{} \omega^V~=~d^{\,V} \tilde{f}~,
\end{equation}
where $\tilde{X}$ is a multivector field on $\vec{J}\upoast E$; this
relation  is of course completely analogous to our equation~(\ref{eq:GHMVF}%
/\ref{eq:HFORM}). Moreover, Kanatchikov introduces a Poisson bracket between
Hamiltonian forms $\tilde{f}$ of degree $n\!-\!r$ and $\tilde{g}$ of degree
$n\!-\!s$, with multivector fields $\tilde{X}$ of degree $r$ and $\tilde{Y}$
of degree $s$ corresponding to $\tilde{f}$ and to $\tilde{g}$ according to
eq.~(\ref{eq:KAN1}), by setting
\begin{equation} \label{eq:KAN2}
 \{\tilde{f},\tilde{g}\}^V~
 =~(-1)^{r(s-1)} \, i_{\tilde{Y}}^{} i_{\tilde{X}}^{} \omega^V~.
\end{equation}
This Poisson bracket satisfies the analogue of the graded Jacobi identity
(\ref{eq:POISBR13}).

We will now show how this approach can be naturally incorporated into the
multisymplectic framework used in the present paper.
\begin{prp}
 Under the canonical projection from extended multiphase space $J\upostar E$
 to ordinary multiphase space $\vec{J}\upoast E$, every Hamiltonian form
 $\tilde{f}$ on $\vec J\upoast E$ as defined by Kanatchikov pulls back to a
 horizontal Poisson form $f$ on $J\upostar E$. Conversely, every horizontal
 Poisson form $f$ of degree $>0$ on $J\upostar E$ is obtained in this way.
 Moreover, the Hamiltonian multivector field $X$ on $J\upostar E$
 corresponding to $f$ can be chosen so as to project to a Hamiltonian
 multivector field $\tilde{X}$ on $\vec{J}\upoast E$ corresponding to
 $\tilde{f}$.
\end{prp}
\textbf{Proof}~~We begin by analyzing the properties of Poisson forms
$f$ of degree $n\!-\!r$ \linebreak ($0<r<n$) on $J\upostar E$ which are
horizontal. Being horizontal, such a form trivially satisfies the kernel
condition (\ref{eq:KERN1}) and its expansion in adapted local coordinates
is
\[
 f~=~\frac{1}{r!} \, f^{\mu_1 \ldots\, \mu_r} \;
     d^{\,n} x_{\mu_1 \ldots\, \mu_r}~,
\]
implying
\begin{eqnarray*}
 d \,\! f \!\!
 &=&\!\! \frac{1}{(r\!-\!1)!} \,
         \frac{\partial f^{\mu_2 \ldots\, \mu_r \nu}}{\partial x^\nu}~
         d^{\,n} x_{\mu_2 \ldots\, \mu_r} \; + \;
         \frac{1}{r!} \,
         \frac{\partial f^{\mu_1 \ldots\, \mu_r}}{\partial q^i}~
         dq^i \,\smwedge\; d^{\,n} x_{\mu_1 \ldots\, \mu_r}                  \\
 & &\!   \mbox{} + \; \frac{1}{r!} \,
         \frac{\partial f^{\mu_1 \ldots\, \mu_r}}{\partial p\>\!_k^\kappa}~
         dp\>\!_k^\kappa \,\smwedge\; d^{\,n} x_{\mu_1 \ldots\, \mu_r} \; + \;
         \frac{1}{r!} \,
         \frac{\partial f^{\mu_1 \ldots\, \mu_r}}{\partial p}~
         dp \;\smwedge\; d^{\,n} x_{\mu_1 \ldots\, \mu_r}~.
\end{eqnarray*}
Comparing this formula with eq.~(\ref{eq:HAMMVF2}), we see that $f$ being
a Hamiltonian form implies first of all that $X$ must be $1$-vertical since
the coefficients $X^{\mu_1 \ldots\, \mu_r}$ give a contribution to $i_X^{}
\omega$ proportional to $\; dq^i \,\smwedge\; dp\;\!_i^\mu \,\smwedge\;
d^{\,n} x_{\mu \mu_1 \ldots\, \mu_r}^{} \;$ which is absent from $df$.
But this implies that $i_X^{} \omega$ contains no terms proportional
to  $\, dp \;\smwedge\; d^{\,n} x_{\mu_1 \ldots\, \mu_r}^{} \,$ either
and hence the coefficients $f^{\mu_1 \ldots\, \mu_r}$ cannot depend on
the energy variable $p\,$; the same then goes for all the coefficients
of~$X$. Therefore, $f$ is the pull-back of a horizontal form $\tilde{f}$
on~$\vec{J}\upoast E$ whereas $X$ projects onto a $1$-vertical multi%
vector field $\tilde{X}$ on $\vec{J}\upoast E$ whose expansion in terms
of adapted local coordinates is given by the second and third term
in eq.~(\ref{eq:HAMMVF1}). Finally, we see that with these relations
between the various objects involved, eq.~(\ref{eq:GHMVF}/\ref{eq:HFORM})
becomes equivalent to eq.~(\ref{eq:KAN1}) plus the relation
\[
 X_0^{\mu_2 \ldots\, \mu_r}~
 =~- \, \frac{\partial f_{\phantom{i}}^{\mu_2 \ldots\, \mu_r \nu}}
             {\partial x_{\phantom{i}}^\nu}~,
\]
which has no counterpart in $\vec{J}\upoast E$ but also does not convey 
any additional information.
\PCPqed

Finally, the fact that the Poisson bracket (\ref{eq:KAN2}) introduced by
Kanatchikov, when pulled back from $\vec{J}\upoast E$ to $J\upostar E$,
coincides with the Poisson bracket defined by eq.~(\ref{eq:POISB2})
follows from the following simple observation.
\begin{prp}
 Let $f$ and $g$ be two horizontal Poisson forms on $J\upostar E$
 of respective degrees $n\!-\!r$ and $n\!-\!s$, with corresponding
 $1$-vertical Hamiltonian multivector fields $X$ and $Y$ of respective
 degrees $r$ and $s$. Then the definition (\ref{eq:POISB2}) of their
 Poisson bracket reduces to the pull-back of eq.~(\ref{eq:KAN2}):
 \begin{equation} \label{eq:POISB4}
  \{f,g\}~=~ =~(-1)^{r(s-1)} \, i_Y^{} i_X^{} \omega~.
 \end{equation}
\end{prp}
\textbf{Proof}~~As we have seen in the proof of the preceding proposition,
$f$ and $g$ being horizontal forces $X$ and $Y$ to be $1$-vertical, so
$i_Y^{} f$ and $i_X^{} g$ vanish. Similarly, eq.~(\ref{eq:HAMMVF3}) shows
that $i_X^{} \theta$ and $i_Y^{} \theta$ are horizontal, so $i_Y^{} i_X^{}
\theta$ and $i_X^{} i_Y^{} \theta$ vanish. Therefore, the exact correction
term of eq.~(\ref{eq:POISB2}) does not contribute in this case. Finally,
$X \smwedge\, Y$ will be $2$-vertical, so contraction of the pull-back
of $\omega^V$ or of $\omega$ with $X$ and $Y$ gives the same result,
implying that eq.~(\ref{eq:POISB4}) is really the pull-back of eq.~%
(\ref{eq:KAN2}).
\PCPqed

\noindent
\textbf{Remark}~~In the case of horizontal Poisson forms, one can also
introduce an associative product, which has been found by Kanatchikov:
\begin{equation}
 f \bullet g~=~\ast^{-1} \left( \ast f \;\smwedge \ast g \right)\!~,
\end{equation}
where $\ast$ is the Hodge star operator on $M$ associated to some metric
which can be transported to horizontal forms on $J\upoast E$ in an obvious
manner. With respect to this product, the Poisson bracket (\ref{eq:POISB4})
satisfies a graded Leibniz rule
\begin{equation}
 \{f,g \bullet h\}~=~\{f,g\} \bullet h \, + \,
                      (-1)^{(r-1)s} \, g \bullet \{f,h\}~.
\end{equation}
However, this product cannot be extended in any natural way to arbitrary
Poisson forms. To see this, suppose we had such an extension at hand. Then
we could define a space of vertical covectors at every point of $J\upostar E$
by requiring it to consist of all covectors that vanish when multiplied by a
horizontal $(n\!-\!1)$-form, which would be equivalent to the choice of a
connection.


\begin{appendix}

\section{Multivector calculus on manifolds}

The extension of the usual calculus on manifolds from vector fields to
multivector fields is by now well known, although it does not seem to be
treated in any of the standard textbooks on the subject. Moreover, there
is a certain amount of ambiguity concerning sign conventions. Our sign
conventions follow those of Tulczyjew \cite{Tul}, but for the sake of
completeness we shall briefly expose the structural properties that
naturally motivate these choices.

Multivector fields of degree $r$ on a manifold are sections of the $r$-th
exterior power of its tangent bundle: they are the dual objects to differential
forms of degree $r$, which are sections of the $r$-th exterior power of its
cotangent bundle. Every known natural operation involving vector fields,
such as the contraction on differential forms, the Lie bracket and the
Lie derivative, has a natural extension to multivector fields: this is
the subject of an area of differential geometry that we simply refer to
as ``multivector \linebreak calculus''. The most important and the ones
that we need in this paper are a) the Schouten bracket between multivector
fields, b) the contraction of a differential form with a multivector field
and c) the Lie derivative of a differential form along a multivector field.

Throughout this appendix, let $M$ be an $n$-dimensional manifold,
$\mathfrak{F}(M)$ the commutative algebra of functions on $M$ (with
respect to pointwise multiplication), $\mathfrak{X}(M)$ the space of
vector fields on $M$ and
\[
 \mathfrak{X}^\wedge(M)~
 =~\bigoplus_{r=0}^n \; \bwedge^{\raisebox{-0.2ex}{${\scriptstyle r}$}}
                        \mathfrak{X}(M)
\]
the supercommutative superalgebra of multivector fields on $M$ (with
respect to pointwise exterior multiplication). 

\subsection{The Schouten bracket}

The Schouten bracket between multivector fields constitutes the natural,
canonical extension both of the Lie bracket between vector fields and of
the Lie derivative of multivector fields (as special tensor fields) along
vector fields. Starting from the Lie derivative of multivector fields along
vector fields, it can be defined by imposing a Leibniz rule with respect to
the exterior product of multivector fields, as in eq.~(\ref{eq:SCHBRLR1})
below.
\begin{prp}
 There exists a unique $\mathbb{R}$-bilinear map
 \begin{equation} \label{eq:SCHBRMP}
  [\,.\,,.\,]~:~\mathfrak{X}^\wedge(M) \times \mathfrak{X}^\wedge(M)~
  \longrightarrow~\mathfrak{X}^\wedge(M)
 \end{equation}
 called the \textbf{Schouten bracket}, with the following properties.
 \begin{enumerate}
  \item It is homogeneous of degree $-1$ with respect to the standard
        tensor degree, i.e.,
        \begin{equation} \label{eq:SCHBRHO}
         \deg X~=~r~~,~~\deg Y~=~s \quad \Longrightarrow \quad
         \deg \, [X,Y]~=~r+s-1~.
        \end{equation}
  \item It is graded antisymmetric: if\/ $X$ has tensor degree~$r$ and\/
        $Y$ has tensor degree~$s$, then
        \begin{equation} \label{eq:SCHBRAS}
         [\>\!Y,X]~=~- \, (-1)^{(r-1)(s-1)} \, [X,Y]~.
        \end{equation}
  \item It coincides with the standard Lie bracket on vector fields.
  \item It satisfies the graded Leibniz rule: if\/ $X$ has tensor degree~$r$,
        $Y$ has tensor degree~$s$ and $Z$ has tensor degree~$t$, then
        \begin{equation} \label{eq:SCHBRLR1}
         [X,Y \smwedge\, Z\>\!]~=~[X,Y] \;\smwedge\; Z \, + \,
                                  (-1)^{(r-1)s} \, Y \>\!\smwedge\; [X,Z\>\!]~.
        \end{equation}
  \item It satisfies the graded Jacobi identity: if\/ $X$ has tensor
        degree~$r$, $Y$ has tensor degree~$s$ and $Z$ has tensor
        degree~$t$, then
        \begin{equation} \label{eq:SCHBRJI}
         (-1)^{(r-1)(t-1)} \, [X,[Y,Z]] \; + \; \mathrm{cyclic~perm.}~=~0~.
        \end{equation}
  \end{enumerate}
\end{prp}
We shall not prove this proposition here but just point out that uniqueness
of an \mbox{operation} with the properties stipulated above follows from the
required $\mathbb{R}$-bilinearity (not $\mathfrak{F}(M)$-bilinearity, of
course), the homogeneity~(\ref{eq:SCHBRHO}), the graded antisymmetry~%
(\ref{eq:SCHBRAS}) and the graded Leibniz rule~(\ref{eq:SCHBRLR1}) alone;
existence can then be proved, for example, by showing that the resulting
local coordinate formula satisfies all these requirements.  Moreover, the
validity of the graded Jacobi identity~(\ref{eq:SCHBRJI}) can be derived
from the standard Jacobi identity for the Lie bracket of vector fields by
means of the graded Leibniz rule (\ref{eq:SCHBRLR1}), using induction on
the degree.

An explicit formula which is slightly more general than the local coordinate
formula just mentioned and often useful in practical applications is that for
the Schouten bracket between decomposable multivector fields; it follows
directly from the same kind of argument and states that for any $r+s$
vector fields $\, X_1,\ldots,X_r \,$ and $\, Y_1,\ldots,Y_s$, we have
\begin{eqnarray} \label{eq:SCHBRDC}
\lefteqn{[ \, X_1 \,\smwedge \ldots \smwedge\, X_r \,,\,
              Y_1 \,\smwedge \ldots \smwedge\, Y_s \, ]}
                                                      \hspace{5mm} \nonumber \\
 &=&\!\! \sum_{i=1}^r \sum_{j=1}^s \, (-1)^{i+j} \; [X_i,Y_j]
         \;\smwedge\; X_1 \,\smwedge \ldots \smwedge\, X_{i-1} \,\smwedge\,
                      X_{i+1} \,\smwedge \ldots \smwedge\, X_r         \\[-3ex]
 & &\!\! \phantom{\sum_{i=1}^r \sum_{j=1}^s \, (-1)^{i+j} \;
         [X_i,Y_j]}
         \;\smwedge\;\> Y_1 \;\smwedge \ldots \smwedge\; Y_{j-1} \,\smwedge\;
                        Y_{j+1} \,\smwedge \ldots \smwedge\; Y_s~.~~\nonumber
\end{eqnarray}
Note also that there is a graded Leibniz rule in the other factor as well:
it follows from the one written down above by using graded antisymmetry
and reads
\begin{equation} \label{eq:SCHBRLR2}
 [X \smwedge\>\! Y,Z\>\!]~=~(-1)^{(t-1)s} \, [Z,X] \;\smwedge\; Y \, + \,
                            X \>\!\smwedge\; [Y,Z\>\!]~.
\end{equation}

Finally, a word seems in order on the adequate choice of signs and degrees.
Indeed, one recognizes eqs~(\ref{eq:SCHBRHO}), (\ref{eq:SCHBRAS}) and~%
(\ref{eq:SCHBRJI}) as the graded homogeneity, the graded antisymmetry
and the graded Jacobi identity familiar from the definition of a Lie
superalgebra, provided one assigns to every multivector field $X$ of
tensor degree $r$ the parity $(-1)^{r-1}\,$: this means that $X$ is even
with respect to the Schouten bracket if it has odd tensor degree and is
odd with respect to the Schouten bracket if it has even tensor degree!
This switch can be better understood by realizing that the operator
$\, \mathrm{ad}(X) = [X,.\,] \,$ lowers the tensor degree of any multi%
vector field that it operates on by $r-1$. The same argument explains
the sign that appears in the graded Leibniz identity~(\ref{eq:SCHBRLR1}),
which can be thought of as stating that the  operator $\, \mathrm{ad}(X)
= [X,.\,] \,$ should be a superderivation with respect to the exterior
product and, more precisely, an even or odd superderivation according
to whether $X$ is even or odd with respect to the Schouten bracket.
We can also think of this operator as defining the Lie derivative
$L_X^{}$ of multivector fields along $X$ \linebreak (possibly up
to signs, which are a matter of convention), but this will not be
needed here.

Algebraically, the situation can be summarized by stating that
$\mathfrak{X}^\wedge(M)$ is a \emph{\mbox{Poisson} superalgebra},
the supersymmetric analogue of a Poisson algebra~-- the structure
encountered, for example, on the space of functions on a symplectic
manifold or, more generally, a Poisson manifold. The surprising aspect
is that this intricate structure requires \emph{no} additional structure
whatsoever on the underlying manifold.

\subsection{Lie derivative of differential forms along multivector fields}

We now come to the other two operations of multivector calculus mentioned
at the beginning of this appendix, namely the contraction of differential
forms with multivector fields and the Lie derivative of differential forms
along multivector fields.

The case of contraction is easy. First, the contraction of a
differential form $\alpha$ with a decomposable multivector field
$\, X_1 \,\smwedge \ldots \smwedge\, X_r \,$ is simply defined as
repeated contraction with its constituents (which by convention
should be performed in the opposite order):
\begin{equation} \label{eq:CONTR1}
 i_{X_1 \wedge \ldots \wedge X_r}^{} \alpha~
 =~i_{X_r}^{} \ldots\, i_{X_1}^{} \alpha~.
\end{equation}
This is then extended to arbitrary (non-decomposable) multivector fields $X$ 
by $\mathfrak{F}(M)$-linearity. (Here, of course, one uses that contraction
is a purely algebraic operation; it would not work so naively if we were
dealing with a differential operator.)

The Lie derivative $L_X^{} \alpha$ of a differential form $\alpha$ along a
multivector field $X$ is most conveniently defined by a generalization of a
well known formula for vector fields.
\begin{dfn} \label{dfn-LieDer}
 On differential forms, the Lie derivative $L_X^{}$ along a multivector
 field $X$ is defined as the supercommutator of the exterior derivative
 $d$ and the contraction operator $i_X^{}$:
 \begin{equation} \label{eq:LDFMVF}
  L_X^{} \alpha~=~d \, i_X^{} \alpha \, - \,
                  (-1)^r \, i_X^{} \, d \>\! \alpha.
 \end{equation}
\end{dfn}
According to the rules of supersymmetry, the sign of the second term is
fixed by observing that $d$ is an odd operator (it is of degree $1$ since
it raises the tensor degree of forms by $1$) while $i_X^{}$ is an even/odd
operator if $r$ is even/odd (it is of degree $-r$ since it lowers the tensor
degree of forms by $r$).
\begin{prp}
 Given any two multivector fields $X$ of tensor degree $r$ and $Y$ of tensor
 degree $s$, we have for any differential form $\alpha$
 \begin{equation} \label{eq:dLX-LXd}
  d \;\! L_X^{} \alpha~=~(-1)^{r-1} \, L_X^{} \;\! d \>\! \alpha~,
 \end{equation}
 \begin{equation} \label{eq:LXiY-iYLX}
  i_{[X,Y]}^{} \alpha~=~(-1)^{(r-1)s} \, L_X^{} i_Y^{} \alpha \, - \,
                        i_Y^{} L_X^{} \alpha~.
 \vspace{1mm}
 \end{equation}
 \begin{equation} \label{eq:LXLY-LYLX}
  L_{[X,Y]}^{} \alpha~=~(-1)^{(r-1)(s-1)} \, L_X^{} L_Y^{} \alpha \, - \,
                        L_Y^{} L_X^{} \alpha~.
 \vspace{1mm}
 \end{equation}
 \begin{equation} \label{eq:iYLX-LYiX}
  L_{X \wedge\>\! Y}^{} \alpha~=~(-1)^s \, i_Y^{} L_X^{} \alpha \, + \,
                                 L_Y^{} i_X^{} \alpha~.
  \end{equation}
\end{prp}
\textbf{Proof}~~The first formula is an immediate consequence of the definition
(\ref{eq:LDFMVF}), since $\, d^{\,2} = 0$. Next, the last formula can be proved
by direct calculation:
\begin{eqnarray*}
 L_{X \wedge\>\! Y}^{} \alpha \!\!
 &=&\!\! d \left( i_{X \wedge\>\! Y}^{} \alpha \right) \, - \,
         (-1)^{r+s} \, i_{X \wedge\>\! Y}^{} \, d \>\! \alpha           \\[1mm]
 &=&\!\! d \left( i_Y^{} i_X^{} \alpha \right) \, - \,
         (-1)^{r+s} \, i_Y^{} i_X^{} \, d \>\! \alpha                   \\[1mm]
 &=&\!\! d \left( i_Y^{} i_X^{} \alpha \right) \, - \,
         (-1)^s \, i_Y^{} \, d \left( i_X^{} \alpha \right)             \\
 & &\!\! \mbox{} + \, (-1)^s \, i_Y^{} \, d \left( i_X^{} \alpha \right) \,
                 - \, (-1)^{r+s} \, i_Y^{} i_X^{} \, d \>\! \alpha      \\[1mm]
 &=&\!\! L_Y^{} i_X^{} \alpha \, + \, (-1)^s \, i_Y^{} L_X^{} \alpha~.
\end{eqnarray*}
Next, observe that the first formula is well known to be true when $X$ and $Y$
are vector fields. The general case follows by induction on the tensor degree
of both factors. Indeed, if $X$, $Y$ and $Z$ are multivector fields of tensor
degree $r$, $s$ and $t$, respectively, such that the above equation holds
for $[X,Y]$ and for $[X,Z]$, one can use the graded Leibniz rule (\ref%
{eq:SCHBRLR1}) to derive that it also holds for $[X,Y \smwedge\, Z\>\!]$:
\begin{eqnarray*}
 i_{[X,Y \wedge Z]}^{} \alpha
 &=&\!\! i_{[X,Y] \wedge Z}^{} \alpha \, + \,
         (-1)^{(r-1)s} \, i_{Y \smwedge [X,Z]}^{} \alpha                \\[1mm]
 &=&\!\! i_Z^{} i_{[X,Y]}^{} \alpha \, + \,
         (-1)^{(r-1)s} \, i_{[X,Z]}^{} i_Y^{} \alpha                    \\[1mm]
 &=&\!\! (-1)^{(r-1)s} \, i_Z^{} L_X^{} i_Y^{} \alpha \, - \,
         i_Z^{} i_Y^{} L_X^{} \alpha                                    \\[1mm]
 & &\!\! \mbox{} + \, (-1)^{(r-1)s+(r-1)t} \, L_X^{} i_Z^{} i_Y^{} \alpha \,
                 - \, (-1)^{(r-1)s} i_Z^{} L_X^{} i_Y^{} \alpha         \\[1mm]
 &=&\!\! (-1)^{(r-1)(s+t)} \, L_X^{} i_{Y \wedge Z}^{} \alpha \, - \,
         i_{Y \wedge Z}^{} L_X^{} \alpha~.
\end{eqnarray*}
Similarly, if $X$, $Y$ and $Z$ are multivector fields of tensor degree $r$,
$s$ and $t$, respectively, such that the above equation holds for $[X,Z]$
and for $[Y,Z]$, one can use the graded Leibniz rule (\ref{eq:SCHBRLR2})
together with eq.~(\ref{eq:iYLX-LYiX}) to derive that it also holds for
$[X \smwedge\>\! Y,Z\>\!]$:
\begin{eqnarray*}
 i_{[X \wedge Y,Z]}^{} \alpha \!\!
 &=&\!\! (-1)^{(t-1)s} \, i_{[X,Z] \wedge Y}^{} \alpha \, + \,
         i_{X \wedge [Y,Z]}^{} \alpha                                   \\[1mm]
 &=&\!\! (-1)^{(t-1)s} \, i_Y^{} i_{[X,Z]}^{} \alpha \, + \,
         i_{[Y,Z]}^{} i_X^{} \alpha                                     \\[1mm]
 &=&\!\! (-1)^{(t-1)s+(r-1)t} \, i_Y^{} L_X^{} i_Z^{} \alpha \, - \,
         (-1)^{(t-1)s} \, i_Y^{} i_Z^{} L_X^{} \alpha                   \\
 & &\!\! \mbox{} + \, (-1)^{(s-1)t} \, L_Y^{} i_Z^{} i_X^{} \alpha \,
                 - \, i_Z^{} L_Y^{} i_X^{} \alpha                       \\[1mm]
 &=&\!\! (-1)^{(r+s-1)t+s} \, i_Y^{} L_X^{} i_Z^{} \alpha \, - \,
         (-1)^s \, i_Z^{} i_Y^{} L_X^{} \alpha                          \\
 & &\!\! \mbox{} + \, (-1)^{(r+s-1)t} \, L_Y^{} i_X^{} i_Z^{} \alpha \,
                 - \, i_Z^{} L_Y^{} i_X^{} \alpha                       \\[1mm]
 &=&\!\! (-1)^{(r+s-1)t} \, L_{X \wedge Y}^{}  i_Z^{} \alpha \, - \,
         i_Z^{} L_{X \wedge Y}^{} \alpha~.
\end{eqnarray*}
Finally, the second formula can now again be proved by direct calculation:
\begin{eqnarray*}
 L_{[X,Y]}^{} \alpha \!\!
 &=&\!\! d \, i_{[X,Y]}^{} \alpha \, + \,
         (-1)^{r+s} \, i_{[X,Y]}^{} \;\! d \>\! \alpha                  \\[1mm]
 &=&\!\! (-1)^{(r-1)s} \, d \>\! L_X^{} i_Y^{} \alpha \, - \,
         d \;\! i_Y^{} L_X^{} \alpha                                    \\
 & &\!   \mbox{} + \, (-1)^{r(s-1)} \, L_X^{} i_Y^{} \;\! d \>\! \alpha \,
                 - \, (-1)^{r+s} \, i_Y^{} L_X^{} \;\! d \>\! \alpha    \\[1mm]
 &=&\!\! (-1)^{(r-1)(s-1)} \, L_X^{} \;\! d \;\! i_Y^{} \alpha \, - \, 
         d \;\! i_Y^{} L_X^{} \alpha                                    \\
 & &\!   \mbox{} + \, (-1)^{r(s-1)} \, L_X^{} i_Y^{} \;\! d \>\! \alpha \,
                 + \, (-1)^s \, i_Y^{} \;\! d \>\! L_X^{} \alpha        \\[1mm]
 &=&\!\! (-1)^{(r-1)(s-1)} \, L_X^{} L_Y^{} \alpha \, - \,
         L_Y^{} L_X^{} \alpha~.
\end{eqnarray*}
\PCPqed

\section{Induced connections}

In this appendix we want to describe briefly the construction of various
induced connections in jet bundle language.

First of all, if $E$ is a fiber bundle over~$M$, we shall view a connection
in~$E$ as a section $\Gamma_E$ of the first order jet bundle~$JE$ of~$E$,
considered as an affine bundle over~$E$; see \cite[Ch.\ IV.17]{KMS}. In
adapted local coordinates $(x^\mu,q^i)$ for~$E$ and $(x^\mu,q^i,q_\mu^i)$
for~$JE$, this section is given by
\[
 \Gamma_E^{} : (x^\mu,q^i)~\mapsto~(x^\mu,q^i,\Gamma_\mu^i(x,q))~.
\]
Next, if $V$ is a vector bundle over~$M$, a linear connection in~$V$ is
given by a section $\Gamma_V$ of $JV$ over~$V$ that depends linearly on
the fiber coordinates. In adapted local coordinates $(x^\mu,v^i)$ for~$V$
and $(x^\mu,v^i,v_\mu^i)$ for~$JV$, this section is given by
\[
 \Gamma_V^{} : (x^\mu,v^i)~\mapsto~(x^\mu,v^i,\Gamma_{\mu j}^i(x) \, v^j)~,
\]
where the $\Gamma_{\mu,j}^i$ are of course the connection coefficients
(gauge potentials) associated with the corresponding covariant derivative.
In particular, a linear connection in the tangent bundle~$TM$ of the base
manifold~$M$ corresponds to a section $\Gamma_{TM}^{}$ of $J(TM)$ over~$TM$
which, in adapted local coordinates $(x^\mu,\dot{x}^\kappa)$ for~$TM$ and
$(x^\mu,\dot{x}^\kappa,\dot{x}_\mu^\kappa)$ for~$J(TM)$ is given by
\[
 \Gamma_{TM}^{} :
 (x^\mu,\dot{x}^\kappa)~\mapsto~
 (x^\mu,\dot{x}^\kappa,\Gamma_{\mu\lambda}^\kappa(x) \, \dot{x}^\lambda)~,
\]
where the $\Gamma_{\mu\lambda}^\kappa$ are of course the corresponding
Christoffel symbols.

Now given a fiber bundle $E$ over $M$ together with a connection in $E$
and a linear connection in $TM$, we can introduce induced connections in
all the various induced bundles that appear in this paper~-- regarded as
fiber bundles over $M$, not over $E$. (This means that jets of sections
will contain just one additional lower space-time index for counting
partial derivatives with respect to the space-time variables.) The
simplest way to describe them is by introducing adapted local coordinates
$(x^\mu,q^i)$ for~$E$ as before; then the local coefficient functions
of the induced connections with respect to the induced adapted local
coordinates can be expressed directly in terms of the local coefficient
functions $\Gamma_\mu^i$ and $\Gamma_{\mu\lambda}^\kappa$ of the original
two connections with respect to the original adapted local coordinates,
as follows.
\begin{itemize}
 \item The vertical bundle $VE$ of $E$:
       \\[1mm]
       in adapted local coordinates
       $(x^\mu,q^i,\dot{q}^k)$ for $VE$ and
       $(x^\mu,q^i,\dot{q}^k,q_\mu^i,\dot{q}_\mu^k)$ \\
       for $J(V^\ast E)$, the induced connection maps
       $(x^\mu,q^i,\dot{q}^k)$ to
       \[
        (x^\mu,q^i,\dot{q}^k,\Gamma_\mu^i(x,q),
         \partial_{\,l}^{} \Gamma_\mu^k(x,q) \, \dot{q}^l)~.
       \]
 \item The dual vertical bundle $V^\ast E$ of $E$:
       \\[1mm]
       in adapted local coordinates
       $(x^\mu,q^i,p\>\!_k^{})$ for $V^\ast E$ and 
       $(x^\mu,q^i,p\>\!_k^{},q_\mu^i,p\>\!_{\mu,k}^{})$ \\
       for $J(V^\ast E)$, the induced connection maps
       $(x^\mu,q^i,p\>\!_k^{})$ to
       \[
        (x^\mu,q^i,p\>\!_k^{},\Gamma_\mu^i(x,q),
         - \, \partial_k^{} \Gamma_\mu^l(x,q) \, p\>\!_l^{})~.
       \]
 \item The pull-back $\pi^*(TM)$ of the tangent bundle $TM$ of $M$ to $E$:
       \\[1mm]
       in adapted local coordinates
       $(x^\mu,q^i,\dot{x}^\kappa)$ for $\pi^*(TM)$ and
       $(x^\mu,q^i,\dot{x}^\kappa,q_\mu^i,\dot{x}_\mu^\kappa)$ \\
       for $J(\pi^*(TM))$, the induced connection maps
       $(x^\mu,q^i,\dot{x}^\kappa)$ to
       \[
        (x^\mu,q^i,\dot{x}^\kappa,\Gamma_\mu^i(x,q),
         \Gamma_{\mu\lambda}^\kappa(x) \, \dot{x}^\lambda)~.
       \]
 \item The pull-back $\pi^*(T^* M)$ of the cotangent bundle
       $T^* M$ of $M$ to $E$:
       \\[1mm]
       in adapted local coordinates
       $(x^\mu,q^i,\alpha_\kappa)$ for $\pi^*(T^* M)$ and
       $(x^\mu,q^i,\alpha_\kappa,q_\mu^i,\alpha_{\mu,\kappa})$ \\
       for $J(\pi^*(T^* M))$, the induced connection maps
       $(x^\mu,q^i,\alpha_\kappa)$ to
       \[
        (x^\mu,q^i,\alpha_\kappa,\Gamma_\mu^i(x,q),
         - \, \Gamma_{\mu\kappa}^\lambda(x) \, \alpha_\lambda)~.
       \]
 \item The pull-back $\pi^*(\bwedge^{\raisebox{-0.2ex}{${\scriptstyle n}$}}\,
       T^* M)$ of the bundle $\bwedge^{\raisebox{-0.2ex}{${\scriptstyle n}$}}\,
       T^* M$ of volume forms on $M$ to $E$:
       \\[1mm]
       in adapted local coordinates
       $(x^\mu,q^i,\epsilon)$ for
       $\pi^*(\bwedge^{\raisebox{-0.2ex}{${\scriptstyle n}$}}\, T^* M)$ and
       $(x^\mu,q^i,\epsilon,q_\mu^i,\epsilon_\mu)$ \\
       for $J(\pi^*(\bwedge^{\raisebox{-0.2ex}{${\scriptstyle n}$}}\, T^* M))$,
       the induced connection maps
       $(x^\mu,q^i,\epsilon)$ to
       \[
        (x^\mu,q^i,\alpha_\kappa,\Gamma_\mu^i(x,q),
         - \, \Gamma_{\mu\rho}^\rho(x) \, \epsilon)~.
       \]
 \item The linearized jet bundle $\vec{J} E$ of $E$:
       \\[1mm]
       in adapted local coordinates
       $(x^\mu,q^i,\vec{q}_\kappa^{\;k})$ for $\vec{J} E$ and
       $(x^\mu,q^i,\vec{q}_\kappa^{\;k},q_\mu^i,\vec{q}_{\mu,\kappa}^{\;k})$ \\
       for $J(\vec{J} E)$, the induced connection maps
       $(x^\mu,q^i,\vec{q}_\kappa^{\;k})$ to
       \[
        (x^\mu,q^i,\vec{q}_\kappa^{\;k},\Gamma_\mu^i(x,q),
         \partial_{\,l}^{} \Gamma_\mu^k(x,q) \, \vec{q}_\kappa^{\;l} \, - \,
         \Gamma_{\mu\kappa}^\lambda(x) \, \vec{q}_\lambda^{\;k})~.
       \]
 \item The jet bundle $JE$ of $E$:
       \\[1mm]
       in adapted local coordinates
       $(x^\mu,q^i,q_\kappa^k)$ for $JE$ and
       $(x^\mu,q^i,q_\kappa^k,q_\mu^i,q_{\mu,\kappa}^k)$ \\
       for $J(JE)$, the induced connection maps
       $(x^\mu,q^i,q_\kappa^k)$ to
       \[
        (x^\mu,q^i,q_\kappa^k,\Gamma_\mu^i(x,q),
         \partial_{\,l}^{} \Gamma_\mu^k(x,q) \,
         (q_\kappa^l - \Gamma_\kappa^l(x,q)) \, - \,
         \Gamma_{\mu\kappa}^\lambda(x) \,
         (q_\lambda^k - \Gamma_\lambda^k(x,q)))~.
       \]
 \item Ordinary multiphase space $\vec{J}\upoast E$:
       \\[1mm]
       in adapted local coordinates
       $(x^\mu,q^i,p\>\!_k^\kappa)$ for $\vec{J}\upoast E$ and
       $(x^\mu,q^i,p\>\!_k^\kappa,q_\mu^i,p\>\!_{\mu,k}^\kappa)$ \\
       for $J(\vec{J}\upoast E)$, the induced connection maps
       $(x^\mu,q^i,p\>\!_k^\kappa)$ to
       \[
        (x^\mu,q^i,p\>\!_k^\kappa,\Gamma_\mu^i(x,q),
         - \, \partial_k^{} \Gamma_\mu^l(x,q) \, p\>\!_l^\kappa \, + \,
         \Gamma_{\mu\lambda}^\kappa(x) \, p\>\!_k^\lambda \, - \,
         \Gamma_{\mu\rho}^\rho(x) \, p\>\!_k^\kappa)~.
       \]
 \item Extended multiphase space $J\upostar E$:
       \\[1mm]
       in adapted local coordinates
       $(x^\mu,q^i,p\>\!_k^\kappa,p\>\!)$ for $J\upostar E$ and
       $(x^\mu,q^i,p\>\!_k^\kappa,p\>\!,
         q_\mu^i,p\>\!_{\mu,k}^\kappa,p\>\!_\mu)$ \\
       for $J(J\upostar E)$, the induced connection maps
       $(x^\mu,q^i,p\>\!_k^\kappa,p\>\!)$ to
       \begin{eqnarray*}
        &&(x^\mu,q^i,p\>\!_k^\kappa,p\>\!,\Gamma_\mu^i(x,q),
           - \, \partial_k^{} \Gamma_\mu^l(x,q) \, p\>\!_l^\kappa \, + \,
           \Gamma_{\mu\lambda}^\kappa(x) \, p\>\!_k^\lambda \, - \,
           \Gamma_{\mu\rho}^\rho(x) \, p\>\!_k^\kappa, \quad \\
        &&~
           - \, \Gamma_{\mu\rho}^\rho(x) \, p \; -
                \left( \partial_\mu \Gamma_\nu^j(x,q) \, - \,
                       \Gamma_\nu^k(x,q) \,
                       \partial_k^{} \Gamma_\mu^j(x,q) \, - \,
                       \Gamma_{\mu\nu}^\kappa(x) \Gamma_\kappa^j(x,q) \right)
                p\>\!_j^\nu)~.
       \end{eqnarray*}
\end{itemize}

\end{appendix}

\section*{Acknowledgements}

Two of the authors (M.F. and H.R) wish to gratefully acknowledge the financial
support of FAPESP (Funda\c{c}\~ao de Amparo \`a Pesquisa do Estado de S\~ao
Paulo, Brazil) which made this collaboration possible.


\end{document}